\documentclass[draft]{agujournal2018}

\usepackage{apacite}
\usepackage{url} 
\usepackage{lineno}
\usepackage{amssymb}
\usepackage{longtable}
\usepackage{amsmath}
\usepackage[framemethod=default]{mdframed}
\DeclareMathOperator{\sign}{sign}
\draftfalse

\journalname{Space Weather}

\begin{document}

%
%


\title{The Challenge of Machine Learning in Space Weather Nowcasting and Forecasting}

%
%




\authors{E. Camporeale}


\affiliation{}{CIRES, University of Colorado, Boulder, CO, USA}
\affiliation{}{Centrum Wiskunde \& Informatica, Amsterdam, The Netherlands}





\correspondingauthor{E. Camporeale}{enrico.camporeale@colorado.edu}




\begin{keypoints}
\item A review on the use of machine learning in space weather.
\item A gentle introduction to machine learning concepts tailored for the space weather community.
\item A selection of challenges on how to beneficially combine machine learning and space weather. 
\end{keypoints}

%
%


\begin{abstract}
The numerous recent breakthroughs in machine learning make imperative to carefully ponder how the scientific community can benefit from a technology that, although not necessarily new, is today living its golden age.
This { Grand Challenge} review paper is focused on the present and future role of machine learning in space weather. 
The purpose is twofold. On one hand, we will discuss previous works that use machine learning for space weather forecasting, focusing in particular on the few areas that have seen most activity: the forecasting of geomagnetic indices, of relativistic electrons at geosynchronous orbits, of solar flares occurrence, of coronal mass ejection propagation time, and of solar wind speed.
On the other hand, this paper serves as a gentle introduction to the field of machine learning tailored to the space weather community and as a pointer to a number of open challenges that we believe the community should undertake in the next decade. The recurring themes throughout the review are the need to shift our forecasting paradigm to a probabilistic approach focused on the reliable assessment of uncertainties, and the combination of physics-based and machine learning approaches, known as gray-box.
\end{abstract}

%
%

%


%
%
%
%

\section{Artificial Intelligence: is this time for real?}
The history of Artificial Intelligence (AI) has been characterized by an almost cyclical repetition of springs and winters: periods of high, often unjustified, expectations, large investments and hype in the media, followed by times
of disillusionment, pessimism, and cutback in funding. Such a cyclical trend is not atypical for a potentially disruptive technology, and it is very instructive to try to learn lessons from (in)famous AI predictions of the past \citep{armstrong2014}, especially now that the debate about the danger of Artificial \emph {General} Intelligence (AGI, that is AI pushed to the level of human ability) is in full swing  \citep{russell2015, russell2016}. Indeed, it is unfortunate that most of the AI research of the past has been plagued by overconfidence and that many hyperbolic statements had very little scientific basis. Even the initial Dartmouth workshop (1956), credited with the invention of AI, had underestimated the difficulty of understanding language processing.

At the time of writing some experts believe that we are experiencing a new AI spring \citep[e.g.][]{bughin2017, olhede2018}, that possibly started as early as 2010. This might or might not be followed by yet another winter.
Still, many reckon that \emph{this time is different}, for the very simple reason that AI has finally entered industrial production, with several of our everyday technologies being powered by AI algorithms.
In fact, one might not realize that, for instance, most of the time we use an app on our smartphone, we are using a machine learning algorithm. The range of applications is indeed very vast: fraud detection \citep{aleskerov1997}, online product recommendation \citep{pazzani2007, ye2009}, speech recognition \citep{hinton2012}, language translation \citep{cho2014}, image recognition \citep{krizhevsky2012}, journey planning \citep{vanajakshi2007}, and many others.\\

Leaving aside futuristic arguments about when, if ever, robotic systems will replace scientists \citep{hall2013}, I think this is an excellent time to think about AI for a space weather scientist, and to try formulating (hopefully realistic) expectations on what our community can learn from embracing AI in a more systematic way. Other branches of physics have definitely been more responsive to the latest developments in Machine Learning.
Notable examples in our neighbor field of astronomy and astrophysics are the automatic identification of exoplanets from the Kepler catalog \citep{pearson2017, shallue2018, kielty2018}, the analysis of stellar spectra from Gaia \citep{fabbro2017, li2017}, and the detection of gravitational waves in LIGO signals \citep{george2018}.\\

\begin{figure}[ht]
\label{fig:terminator}
\centering
 \includegraphics[width=30pc]{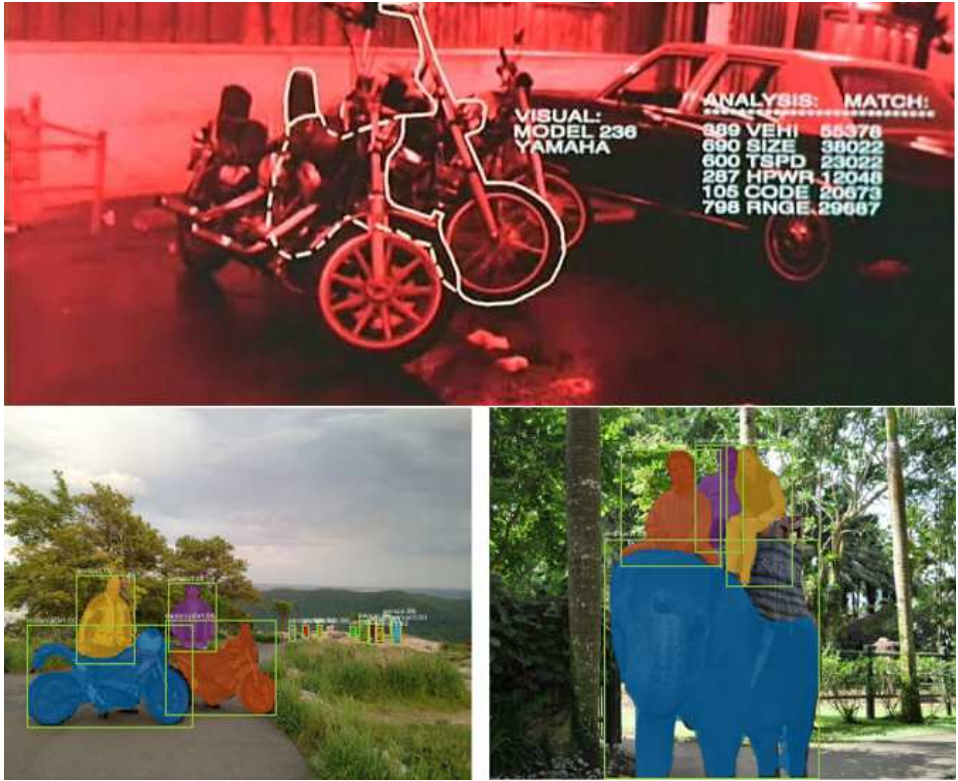}
\caption{Top: scene from the \emph{Terminator 2} movie (1991). Bottom: examples of segmentation problems as solved by Mask R-CNN (2018) \citep{he2017}.}
\end{figure}

Each generation has its own list of science fiction books and movies that have made young kids fantasize about what the future will look like after AGI will finally be achieved. Without digressing too much, I would just like to mention one such iconic movie for my generation, the \emph{Terminator} saga. In the second movie, a scene is shown from the cyborg point of view. The cyborg performs what is today called a segmentation problem, that is identifying single, even partially hidden, objects from a complex image (specifically, the movie's hero is intent in choosing the best motorcycle to steal). The reason I am mentioning this particular scene is that, about 30 years later, a landmark paper has been published showing that solving a segmentation problem is not science fiction anymore (see Figure \ref{fig:terminator}) \citep{he2017}. Not many other technologies can claim to have made fiction come true, and in a such short time frame!

\section{The Machine Learning renaissance}\label{sec:ML_renaissance}
One of the reasons why the current AI spring might be very different from all the previous ones and in fact never revert to a winter is the unique combination of three factors that have never been simultaneously experienced in our history. First, as we all know, we live in the time of big data. The precise meaning of what constitutes big data depends on specific applications. In many fields the data is securely guarded as the gold mine on which a company's wealth is based (even more than proprietary, but imitable, algorithms). Luckily, in the field of Space Weather most of the data and associated software is released to the public.\citep{NAS2018}\\
The second factor is the recent advancement in {Graphics Processing Units} (GPU) computing. In the early 2000s GPU producers (notably, Nvidia) were trying to extend their market to the scientific community by depicting GPUs as accelerators for high performance computing (HPC), hence advocating a shift in parallel computing where CPU clusters would be replaced by heterogeneous, general-purpose, GPU-CPU architectures. Even though many such machines exist today, especially in large HPC labs worldwide, I would think that the typical HPC user has not been persuaded to fully embrace GPU computing (at least in space physics), possibly because of the steep learning curve required to proficiently write GPU codes. More recently, during the last decade, it has become clear that a much larger number of users (with respect to the small niche of HPC experts) was ready to enter the GPU market: machine learning practitioners (along with bitcoin miners!). And this is why GPU companies are now branding themselves as enablers of the machine learning revolution. 

It is certainly true that none of the pioneering advancements in machine learning would have been possible without GPUs. As a figure of merit, the neural network NASnet, that delivers state-of-the-art results on classification tasks of ImageNet and CIFAR-10 datasets, required using 500 GPUs for 4 days (including search of optimal architecture) \citep{zoph2017}.
Hence, a virtuous circle, based on a larger and larger number of users and customers has fueled the faster than Moore's law increase in GPU speed witnessed in the last several years. The largest difference between the two group of GPU users targeted by the industry, i.e. HPC experts and machine learning practitioners (not necessarily experts) is in their learning curve. While a careful design and a deep knowledge of the intricacies of GPU architectures is needed to successfully accelerate an HPC code on GPUs, it is often sufficient to switch a flag for a machine learning code to train on GPUs. 

\begin{figure}[h]
 \centering
 \includegraphics[width=\columnwidth]{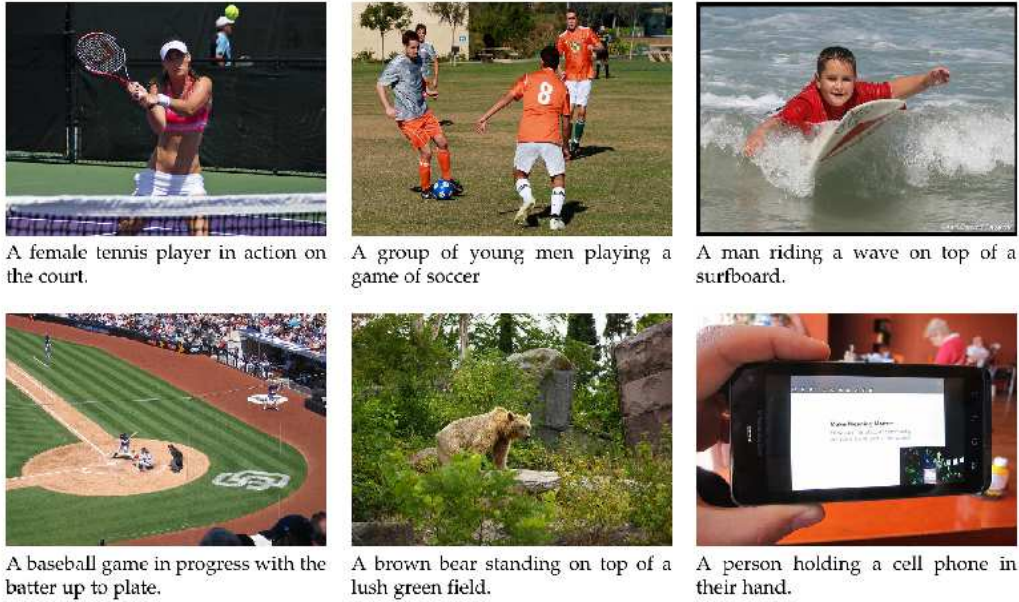}
 \caption{Automatically generating captions to images represents a state-of-the-art achievement in Machine Learning, that combines image recognition and natural language processing. Figure taken from the arXiv version of \citet{donahue2015} (arXiv:1411.4389). }
 \label{fig:auto_caption}
  \end{figure}

This fundamental difference leads us to the third enabling factor of the machine learning renaissance: the huge money investments from IT companies, that has started yet another virtuous circle in software development. Indeed, companies like Google or Facebook own an unmatchable size of data to train their machine learning algorithms. By realizing the profitability of machine learning applications, they have largely contributed to the advancement of machine learning, especially making their own software open-source and relatively easy to use~\citep[see, e.g.,][]{abadi2016}. Arguably, the most successful applications of machine learning are in the field of computer vision. Maybe because image recognition and automatic captioning are tasks that are very easy to understand for the general public, this is the field where large IT companies have advertised their successes to the non-experts and attempted to capitalize them. Classical examples are the Microsoft bot that guesses somebody's age (https://www.how-old.net), that got 50 million users in one week, and the remarkably good captioning bot www.captionbot.ai (see Figure \ref{fig:auto_caption} from \citet{donahue2015} for a state-of-the-art captioning example).

{In a less structured way, the open-source scientific community has also largely contributed to the advancement of machine learning software. Some examples of community-developed python libraries that are now widely used are \emph{theano}~\citep{bergstra2010}, \emph{scikit-learn} \citep{pedregosa2011}, \emph{astroML}~\citep{vanderplas2012}, \emph{emcee}\citep{vanderplas2012}, \emph{PyMC}~\citep{patil2010}, among many others.
This has somehow led to an explosion of open-source software which is very often overlapping in scope. Hence, ironically the large number of open-source machine learning packages available might actually constitute a barrier to somebody that entering the field is overwhelmed by the amount of possible choices. In the field of heliophysics alone, the recent review by \citet{burrell2018} compiles a list of 28 python packages.}

As a result of the unique combination of the three above discussed factors, for the first time in history a layperson can easily access Terabytes of data (big data), afford to have a few thousand cores at {their} disposal (GPU computing), and easily train a machine learning algorithm with absolutely no required knowledge of statistics or computer science (large investments from IT companies in open-source software).\\

The purpose of this review is twofold. On one hand, I will discuss previous works that use machine learning for space weather forecasting. The review will be necessarily incomplete and somewhat biased, and I apologize for any relevant work I might have overlooked. In particular, I will focus on a few areas, where it seems that several attempts of using machine learning have been proposed in the past: the forecasting of geomagnetic indices, of relativistic electrons at geosynchronous orbits, of solar flares occurrence, of coronal mass ejection propagation time, and of solar wind speed.
On the other hand, this paper serves as a gentle introduction to the field of machine learning tailored to the space weather community and, as the title suggests, as a pointer to a number of open challenges that I believe the community should undertake in the next decade. In this respect, the paper is recommended to bold and ambitious PhD students!\\

This review is organized as follows. Section \ref{sec:ML_SW} briefly explains why and how space weather could benefit from the above described machine learning renaissance. It also concisely introduces the several tasks that a machine learning algorithm can tackle, and the appropriate performance metrics for each task. Section \ref{sec_review} constitutes the review part of the paper. Each subsection (geomagnetic indices, relativistic electrons at GEO, solar images) is concluded with a recapitulation and an overview of future perspective in that particular field. Section \ref{sec:new_trends} discusses a few new trends in machine learning that I anticipate will soon see an application in the process of scientific discovery. Section \ref{sec:conclusions} concludes the paper by discussing the future role of machine learning in space weather and space physics, in the upcoming decade, and by commenting my personal selection of open challenges that I encourage the community to consider.

\section{Machine Learning in Space Weather}\label{sec:ML_SW}
How can Space Weather benefit from the ongoing machine learning revolution? First of all, I would like to clarify that space weather is not new to machine learning. As many other subjects that are ultimately focused on making predictions, several attempts to use (mainly, but not only) neural networks have been made since the early 90's. This will be particularly clear in Section \ref{sec_review}, which is devoted to a (selected) review of past literature. Especially in some areas such as geomagnetic index prediction, the list of early works is quite overwhelming. Before proceeding in commenting how machine learning can be embraced by the space weather community, it is therefore necessary to address the (unfortunately still typical) skeptical reaction of many colleagues that wonder `if everything (i.e. any machine learning technique applied to space weather) has been tried already, why do we need to keep trying?' There are two simple answers, in my opinion. First, not \emph{everything} has been tried; in particular, deep learning based on convolutional neural networks (CNN), which incidentally is one of the most successful trend in machine learning \citep{lecun2015}, has been barely touched in this community. Second, machine learning has never been as successful as it is now: this is due to the combination of the three factors discussed in Section \ref{sec:ML_renaissance} thanks to which it is now possible to train and compare a large number of models on a large size dataset. In this respect, it is instructive to realize that the basic algorithm on which a CNN is based has not changed substantially over the last 30 years \citep{lecun1990}. What has changed is the affordable size of a training set, the software (open-source python libraries) and the hardware (GPUs). Hence, this is the right time when it is worth to re-test ideas proposed ten or twenty years ago, because what did not seem to work then might prove very successful now.\\
Space Weather possess all the ingredients often required for a successful machine learning application. As already mentioned, we have a large and freely available dataset of in-situ and remote observations collected over several decades of space missions . Restricting our attention on satellites typically used for space weather predictions, the Advanced Composition Explorer (ACE), Wind and the Deep Space Climate Observatory (DSCOVR) provide in-situ plasma data in proximity of the first Lagrangian point (L1), with several temporal resolution, some of which dates back 20 years. The Solar and Heliospheric Observatory (SOHO) and the Solar Dynamics Observatory (SDO) provide Sun images at different wavelengths, magnetograms and coronographs, also collectively covering a 20 years period. Moreover, the OMNI database collects data at both hour and minutes frequency of plasma and solar quantities, as well as geomagnetic indices. Other sources of space weather data are the twin Van Allen Probes (VAP) whose database is now quite sizable, having entered their 7th year of operation; the Geostationary Operational Environmental Satellite system (GOES)  provides measurements of geomagnetic field, particle fluxes and X-rays irradiance at geostationary orbit. Recently, 16 years of GPS data have been released to the public, providing a wealth of information on particle fluxes \citep{morley2017}. Particle precipitation is measured by the Polar Operational Environmental Satellites (POES) and the Defense Meteorological Satellite Program (DMSP).In addition to space-based measurements, an array of ground-based magnetometer monitors the Earth's magnetic field variation on timescales of seconds. A list of data sources for space weather is in Table \ref{tab:sat}.\\
 
 \begin{table}\label{tab:sat}
 \caption{Data used for Space Weather.}
 \centering
 \begin{tabular}{ll}
 \hline
  ACE & http://www.srl.caltech.edu/ACE/\\
  Wind & https://wind.nasa.gov/\\
  DSCOVR & https://www.nesdis.noaa.gov/content/dscovr-deep-space-climate-observatory\\
  SOHO & https://sohowww.nascom.nasa.gov/\\
  SDO & https://sdo.gsfc.nasa.gov/\\
  OMNI & https://omniweb.gsfc.nasa.gov/index.html\\
  VAP & http://vanallenprobes.jhuapl.edu/\\
  GOES & https://www.goes.noaa.gov\\
  POES& https://www.ospo.noaa.gov/Operations/POES/index.html\\
  GPS &  https://www.ngdc.noaa.gov/stp/space-weather/satellite-data/satellite-systems/gps/\\
  DMSP & https://www.ngdc.noaa.gov\\
  Ground-based magnetometers & http://supermag.jhuapl.edu/\\
  \hline  
 \end{tabular}
 \end{table}

Furthermore, we have rather sophisticated physics-based models and a fair understanding of the physics processes behind most space weather events. The fact that a first-principle approach will never be feasible for forecasting space weather events is essentially due to the large separation of scales in space and time involved, to the short time lag between causes and effects, and the consequent enormous computational cost of physics-based models. In this respect, I believe that it is fair to say that the space weather community has a good understanding of why some models have poor forecasting capabilities (for instance what is the missing physics in approximated models \citep[see,e.g.][]{welling2017}) and what links of the space weather prediction chain will benefit more to a coupling with a data-driven approach. Therefore, space weather seems to be an optimal candidate for a so-called \emph{gray-box} approach. 

As the name suggests, the gray-box paradigm sits in between two opposites approaches. 
{For the purpose of this paper, black-box methods refer to ones that are completely data-driven, seeking empirical correlations between variables of interests, and do not typically use a-priori physical information on the system of interest. Machine learning falls in this category (but see Section \ref{sec:new_trends} for recent trends in machine learning that do make use of physics law). On the other end of the spectrum of predictive methods, white-box models are based on assumptions and equations that are presumed valid, irrespective of data (just in passing I note that physics is an experimental science, therefore most physical laws are actually rooted in data validation. However, once a given theory stands the test of time, its connection to experimental findings is not often questioned or checked). All physics-based models, either first-principle or based on approximations, are white-box. Note that this distinction is different from what the reader can found in other contexts. For instance, in Uncertainty Quantification or in Operations Research, a model is said to be used as a black box whenever the internal specifics are not relevant. Other uses of the white vs black-box paradigm involve the concept of interpretability \citep{molnar2018}. However, I find that concept too subjective to be applied rigorously and dangerously prone to philosophical debates.}
Table (\ref{tab:gray}) succinctly describes the advantages and disadvantages of the two approaches, namely computational speed and ability to generalize to out-of-sample (unseen or rare) data.\\

 \begin{table}\label{tab:gray}
 \caption{Comparison between white- and black-box approaches.}
 \centering
 \begin{tabular}{c|  p{5cm}  p{5cm}}
 \hline
 & White (physics-based) & Black (data-driven)  \\
 \hline
  Computational cost & Generally expensive. Often not possible to run in real-time. & Training might be expensive (depending on the datasize) but execution is typically very fast.\\ 
  Robustness    &Robust to unseen data and rare events.  & Not able to extrapolate outside the range of the training set.\\
  Assumptions & Based on physics approximations. & Minimal set of assumptions.\\
  Consistency with observations & Verified a posteriori. & Enforced a priori.\\
  Steps towards a gray-box approach& Data-driven parameterization of inputs. & Enforcing physics-based constraints.\\
  Uncertainty quantification & Usually not built-in. It requires Monte-Carlo ensemble. & It can be built-in.
 \end{tabular}
 \end{table}

In the gray-box paradigm one tries to maximize the use of available information, be it data or prior physics knowledge. 
{Hence, a gray-box approach applies either when a physics-based model is enhanced by data-derived information, or when a black-box model incorporates some form of physics constraints.}
In the field of space weather there are at least three ways to implement a gray-box approach.
First, by realizing that even state-of-the-art models rely on ad-hoc assumptions and parameterization of physical inputs, one can use observations to estimate such parameters. This usually leads to an inverse problem (often ill-posed) that can be tackled by Bayesian parameter estimation and data assimilation (see, e.g. \citet{reich2015} for an introductory textbook on the topic). Bayes' theorem is 
the central pillar of this approach. It allows to estimate the probability of a given choice of parameters, conditioned on the observed data, as a function of the likelihood that these data are indeed observed when the model uses the chosen parameters. In mathematical terms, Bayes' formula expresses the likelihood of event $A$ occurring when event $B$ is true, $p(A|B)$ as a function of the likelihood of event $B$ occurring when event $A$ is true, $p(B|A)$. In short, the parameters that we seek to estimate are treated as a multidimensional random variable $\mathbf{m}$ (for model), that is related to observations (data) $\mathbf{d}$ through a forward model (the physics-based equations): $F(\mathbf{m})\approx \mathbf{d}$.
The quantity of interest is the so-called posterior probability density function (PPDF) of $\mathbf{m}$, given $\mathbf{d}$, that is
calculated by Bayes' formula:
\begin{linenomath*}
\begin{equation}
p(\mathbf{m}|\mathbf{d}) \propto p(\mathbf{d}|\mathbf{m}) p(\mathbf{m})
\end{equation}
\end{linenomath*}
where $p(\mathbf{d}|\mathbf{m})$ is a conditional probability known as likelihood, and $p(\mathbf{m})$ is called
the prior, and it represents the knowledge (or assumptions) of $\mathbf{m}$ before looking at the data. The computational cost of
this procedure resides in calculating the likelihood which, for instance, can be expressed as
$p(\mathbf{d}|\mathbf{m}) \propto \exp (- || F(\mathbf{m}) - \mathbf{d} || /2\sigma^2)$
and requires to solve the forward model for each given choice of $\mathbf{m}$. The standard procedure, for high dimensional problems (i.e., large number of parameters) is to resolve to a Markov Chain Monte-Carlo (MCMC) approach \citep{kennedy2001, gelman2013}.
However, MCMC requires to run a large ensemble of forward models that are often costly simulations. More efficient methods based on the combination of machine learning, sparse grid collocation, and Monte-Carlo have recently been proposed (see, e.g. \citet{jin2008, ma2009}).\\

A second gray-box approach is the following. Space weather predictions are produced by a chain of interconnected models that solve different physics equations in different parts of the Sun-Earth domain. Loosely speaking, (at least) four domains are studied separately: the Sun surface to the bow shock (solar wind), the magnetosphere, the radiation belt, and the ionosphere (down to ground). In each of these models there are components that might be successfully replaced by a machine learning module, that is by a \emph{surrogate} model that (once trained) has a much lower computational demand and similar accuracy.\\

Finally, for many quantities of interest prediction algorithms have been studied based completely either on a black- or on a white-box approach, that is using either data- or physics-based models. It would be a worthwhile effort to develop ensemble predictions based on a combination of models, where the weights assigned to each model are learned depending, e.g., on geomagnetic conditions. Ensemble modeling has been shown to be very effective in space weather applications \citep{murray2018, morley2018}.\\

Having sketched the general trend and future possibilities of using machine learning in space weather, we now move to a more detailed description of different tasks that can be tackled by machine learning algorithms. This is still a concise description and we refer the reader to specialized textbooks \citep[e.g.,][]{bishop2006, murphy2012} and dedicated monographs \citep{camporeale2018}.
A nomenclature well-established in the machine learning community is to describe a task as supervised or unsupervised, depending whether the user has access to a `ground truth' for the output of interest or not (that is either no ground truth exists, or we do not know what it is). We use the same nomenclature in the following.

\subsection{Supervised regression}\label{sec:regression}
Let us assume that we want to find a nonlinear map between a set of multidimensional inputs $\mathbf{x}=(x_1,x_2,\ldots,x_{N_i})$ and its corresponding scalar output $y$, under the general form 
\begin{equation}\label{def_y}
 y = f(\mathbf{x})+\varepsilon
\end{equation}
where  $f:\mathbb{R}^{N_i}\rightarrow \mathbb{R}$ is a nonlinear function and $\varepsilon$ is a stochastic error (noise) term.
If we have access to a list of observations $\{\mathbf{x}^i_{obs},y^i_{obs}\}$ of size $N_D$, this constitutes a supervised regression problem. Depending on what assumptions we make on the function $f$ and on the error term $\varepsilon$, this problem can be solved by a large variety of methods. All of the methods, however, can be understood as an optimization problem. Indeed, any regression problem can be set up as finding the unknown map $f$ that minimizes a given cost function. In turn, the cost function is defined as a function of the observed values $y^i_{obs}$ and the predictions $\hat{y}^i=f(\mathbf{x}^i_{obs})$, for a certain number of training data $i=1,\ldots,N_T$. Examples of cost functions are the mean squared error $MSE=\frac{1}{N_T}\sum_{i=1}^{N_T}(\hat{y}^i-y^i_{obs})^2$ and the mean absolute error $MAE=\frac{1}{N_T}\sum_{i=1}^{N_T}|\hat{y}^i-y^i_{obs}|$. In practice, the unknown function $f$ is restricted to a given class that is chosen a priori. For instance, the first method we encounter in a statistics textbook is probably linear regression solved by the method of least squares. In that case, $f$ is defined as $f=a\mathbf{x}+b$, with $a$ a row vector of size $N_i$ and $b$ a scalar. The assumption on the error term $\varepsilon$ is that it is normally distributed, and the corresponding cost function is the $MSE$. \\
Note that excluding the error term in the definition (\ref{def_y}) transforms the regression into an interpolation problem. Interpolation is less interesting, because it assumes that a nonlinear function $f$ exists that maps \emph{exactly} $\mathbf{x}$ into $y$. In other words, the term $\varepsilon$ takes into account all possible reasons why such exact mapping might not exist, including observational errors and the existence of latent variables. In particular, different values of $y$ might be associated to the same input $\mathbf{x}$, because other relevant inputs have not been included in $\mathbf{x}$ (typically because not observed, hence the name latent).\\
The input $\mathbf{x}$ and the output $y$ can be taken as quantities observed at the same time, in which case the problem is referred to as \emph{nowcasting}, or with a given time lag, which is the more general \emph{forecasting}. 
In principle a supervised regression task can be successfully set and achieve good performances for any problem for which there is a (physically motivated) reason to infer some time-lagged causality between a set of drivers and an output of interest.   
In general, the dimension of the input variable can be fairly large. For instance, one can employ a time history of a given quantity, recorded with a certain time frequency. Examples of supervised regression in space weather are the forecast of a geomagnetic index, as function of solar wind parameters observed at L1 \citep{gleisner1996, weigel1999, macpherson1995, lundstedt1994, uwamahoro2014, valach2009}, the prediction of solar energetic particles (SEP) \citep{gong2004, li2008, fernandes2015}, of the f10.7 index for radio emissions \citep{huang2009, ban2011}, of ionospheric parameters  \citep{chen2010}, of sunspot numbers or, more in general, of the solar cycle \citep{calvo1995, fessant1996, ashmall1997, conway1998, uwamahoro2009, lantos1998, pesnell12}, of the arrival time of interplanetary shocks \citep{vandegriff2005}, and of coronal mass ejections \citep{sudar2015, choi2012}.\\
Regression problems typically output a single-point estimates as a prediction, lacking any way of estimating the uncertainty associated to the output. Methods exist that produce probabilistic outputs, either by directly using neural networks \citep{gal2016}, or by using Gaussian Processes \citep{rasmussen2004}. More recently, a method has been developed to directly estimate the uncertainty of single-point forecast, producing calibrated Gaussian probabilistic forecast \citep{camporeale2019}.
The archetype method of supervised regression is the neural network (NN). See Box 1 for a short description of how a NN works.\\

\begin{mdframed}[nobreak=true]
\sloppy \footnotesize

{\bf Box 1: Neural Networks: a short tour with some Math and no Biology}\\
        
        A Neural Network (NN) is a powerful and elegant way of approximating a complex nonlinear function as a composition of elementary nonlinear functions. In its simplest form a NN takes a multidimensional input argument $\mathbf{x}=\{x_1,x_2,\ldots,x_{N_i}\}$ of dimension $N_i$ and outputs a single scalar $y$, by applying the following mapping:
        \begin{equation}\label{eq_NN}
         y(\mathbf{x}) = \sum_{i=1}^q w_i\sigma\left( \sum_{j=1}^{N_i} a_{ij} x_j + b_i\right),
        \end{equation}
	where $\sigma(\cdot)$ is a continuous nonlinear function (in jargon called \emph{activation function}). Historically activation functions were chosen as sigmoids, i.e. with $\lim_{s\rightarrow\infty}\sigma(s)=1$ and $\lim_{s\rightarrow-\infty}\sigma(s)=0$. Modern NN use a REctified Linear Unit (RELU) or some modifications of it as an activation function. A RELU $\sigma$ holds $\sigma(s<0)=0$ and $\sigma(s\geq0)=s$. In Eq.(\ref{eq_NN}), $w_i$ and $a_{ij}$ represent weights and $b$ is a so-called bias vector. Effectively $w$, $a$ and $b$ represent free parameters that need to be optimized. A NN represented by Eq. (\ref{eq_NN}) is called a \emph{single hidden-layer feedforward} network. In simple words, the input vector goes first through a linear transformation by the weights $a$ and the bias vector $b$ (this can be represented as a matrix-vector multiplication). The new vector resulting from such transformation is then fed into the activation function. This operation is repeated $q$ times (each time with different weights $a$ and biases $b$), and in turn the $q$ results of $\sigma(\cdot)$ are again linearly combined through the weight vector $w$. The number $q$ is a free parameter, in jargon called \emph{number of neurons}. Eq. (\ref{eq_NN}) might look as a cumbersome mathematical construct, and not an intuitive way of defining an approximation for a given nonlinear function. However, the theory of NN has a strong mathematical foundation, in the proof that Eq. (\ref{eq_NN}) can approximate any continuous function with arbitrary precision, for $q$ large enough \citep{cybenko1989}. A practical way of understanding NN, especially when compared to more traditional methods is that the superposition of activation functions provide a much richer basis function, being optimized (through the fine-tuning of the free parameters) to the nonlinear function that is approximated. An open question remains on how to judiciously choose the values for the weights and biases.
	This is done through training using \emph{backpropagation}. First, a cost function needs to be chosen (see Section \ref{sec:regression}) that measures the distance between the observed and predicted output values. The optimization problem that the NN seeks to solve is to minimize a given cost function. Because Eq.(\ref{eq_NN}) is analytical, one can compute the derivative of the cost function with respect to each weight, by a simple application of the chain rule of differentiation. Once these derivatives are computed, an iterative gradient descent method can be applied.\\
	What is \emph{Deep Learning}? The output of Eq.(\ref{eq_NN}) can be used as an input to another set of activation functions (not necessarily with the same functional form), which then can be fed to yet another layer and so on. In this way one can construct a \emph{multi-layer} neural network by a simple concatenation of single layers. It is said that the network grows in depth, hence Deep Learning. Going back to the basis function interpretation, the advantage of going deep is that the family of functional forms that can be represented becomes much larger, giving the network a larger expressive power. The downside, of course, is that the number of weights also increases and the related training cost and overfitting problems.\\
	What is a \emph{Convolutional Neural Network (CNN)?} The structure described above constitutes what is called a dense layer. When the input is highly-dimensional, like in the case of an image where each pixel represents an input, dense layers can rapidly result in a too-large number of weights. One solution is to replace the matrix-vector multiplication in Eq.(\ref{eq_NN}) to a convolution operation. In this case, a discrete filter (for instance a 3x3 matrix) is introduced and the unknown weights to be optimized are the entries of the filter. The filtered input (that is the image convolved with the filter) is then fed to an activation function, similarly to a standard neural network. Also, a Convolutional layer is often part of a deep network, where the output of a layer is fed as the input of the next layer. 
	By using CNN, there are two advantages. First, the number of weights is reduced and input-independent, with respect to a dense layer network. Second, the application of a filtering operation is particularly well posed when dealing with images. In fact, filtering can extract spatial features at a given characteristic scale, while retaining spatial transformation invariance (such as translation or rotation invariance). Moreover, the repeated application of filters can process the input image on a number of different scales and different level of feature abstraction.
\end{mdframed}

\normalsize
\subsection{Supervised classification}\label{sec:supervised_class}
The question that a supervised classification task answers is: What class does an event belong to? This means that a list of plausible classes has been pre-compiled by the user, along with a list of examples of events belonging to each individual class (supervised learning). This problem is arguably the most popular in the machine learning community, with the ImageNet challenge being its prime example \citep{ILSVRC15, deng2009}. The challenge, that has been active for several years and it is now hosted on the platform kaggle.com, is to classify about hundred thousands images in 1000 different categories. In 2015 the winners of the challenge (using deep neural networks) have claimed to have outperformed human accuracy in the task. \\
In practice any regression problem for a continuous variable can be simplified into a classification task, by introducing arbitrary thresholds and dividing the range of predictands into `classes'. One such example, in the context of Space Weather predictions, is the forecast of solar flare classes. Indeed, the classification into A-, B-, C-, M-, and X- classes is based on the measured peak flux in (W/m$^2$) arbitrarily divided in a logarithmic scale. In the case of a 'coarse-grained' regression problem, the same algorithms used 
for regression can be used, with the only change occurring in the definition of cost functions and a discrete output. 
For instance, a real value output $z$ (as in a standard regression problem) can be interpreted as the probability of the associated event being true or false (in a binary classification setting), by squashing the real value through a so-called logistic function:
\begin{linenomath*}
\begin{equation}
 \hat{y}=\sigma(z) = \frac{1}{1+e^{-z}}.
\end{equation}
\end{linenomath*}
Because $\sigma(z)$ is bounded between 0 and 1, its probabilistic interpretation is straightforward. Then, a simple and effective cost function is the cross-entropy $C$, defined as:
\begin{linenomath*}
\begin{equation}
 C(y,z) = (y-1)\log(1-\sigma(z)) - y\log(\sigma(z))
\end{equation}
\end{linenomath*}
where $y$ is the ground true value of the event (0-false or 1-true) and $z$ is the outcome of the model, squashed in the interval $[0,1]$ via $\sigma(z)$. One can verify that $C(y,z)$ diverges to infinity when $|y -\hat{y}|=1$, that is the event is completely mis-specified, and it tends to zero when $|y -\hat{y}|\rightarrow 0$. {This approach is called logistic regression (even though it is a classification problem).}\\
Other problems represent proper classification tasks (i.e. in a discrete space that is not the result of a coarse-grained discretization of a continuous space). Yet, the underlying mathematical construct is the same. Namely, ones seeks a nonlinear function $f$ that maps multidimensional inputs to a scalar output as in Eq.(\ref{def_y}) and whose predicted values $\hat{y}$ minimize a given cost function.
In the case of image recognition, for instance, the input is constituted by images that are flattened into arrays of pixel values. 
A popular classifier is the Support-Vector Machine \citep{vapnik2013}, that finds the hyperplane that optimally divides the data to be classified (again according to a given cost function) in its high-dimensional space (equal to the dimensionality of the inputs), effectively separating individual events into classes.\\
In the context of Space Weather an example is the automatic classification of sunspot groups according to the McIntosh classification \citep{colak2008}, or the classification of solar wind into types based on different solar origins \citep{camporeale2017}.
It is useful to emphasize that, contrary to regression problems, interpreting the output of a classification task from a probabilistic perspective is much more straightforward, when using a sigmoid function to squash an unbounded real-value output to the interval $[0,1]$. However, some extra steps are often needed to assure that such probabilistic output is well calibrated, that is it is statistically consistent with the observations (see, e.g. \citet{zadrozny2001,niculescu2005}). 

\subsection{Unsupervised classification, a.k.a. clustering}\label{sec:clustering}
Unsupervised classification applies when we want to discover similarities in data, without deciding a priori the division between classes, or in other words without {specifying classes and their labels}. Yet again, this can be achieved by an optimization problem, where the `similarity' between a group of events is encoded into a cost function. This method is well suited in cases when a 'ground truth' cannot be easily specified. This task is harder (and more costly) than supervised classification, since a criterion is often needed to specify the optimal number of classes. A simple and often used algorithm is the so-called k-means, where the user specifies the number of clusters $N_k$, and each observation $\mathbf{x}^i=(x^i_1,x^i_2,\ldots,x^i_{N_i})$ is assigned to a given cluster. The algorithm aims to minimize the within-cluster variance, defined as $\sum_{k=1}^{N_K}\sum_{i\in S_k} ||\mathbf{x}^i - \mathbf{\mu}_k||^2$, where the first sum is over the number of clusters, the second sum is over the points assigned to the cluster $k$, and $\mathbf{\mu}_k$ is the centroid of cluster $k$.\\
An unsupervised neural network is the self-organizing map \citep{kohonen1997}. The output of the network is a two-dimensional topology of neurons, each of which maps to a specific characteristic of the inputs. {In a self-organizing map, similar inputs activate close-by neurons, hence aggregating them into clusters.} Even though some initial choice and constraint in the network architecture need to be done, this method dispenses from choosing a priori the number of clusters and it indeed gives a good indication of what an optimal number might be.\\
In Space Weather, an unsupervised classification of the solar wind has been performed in \citet{heidrich2018}, and a self-organizing map has been applied to radiation belt particle distributions in \citet{souza2018}.
It is fair to say, however, that the majority of past studies have focused on supervised learning.

\subsection{Dimensionality reduction}\label{sec:dim_red}
The last family of methods that we concisely describe is dimensionality reduction. This is a family of techniques that aims at reducing the size of a dataset, preserving its original information content, with respect to a specific prediction objective. It is very important in the context of multidimensional datasets, such as when working with images, since a dataset can easily become very sizable and data handling becomes a major bottleneck in the data science pipeline. A dimensionality reduction technique can be also used to rank the input variables in terms of how important they are with respect to forecasting an output of interest, again with the intent of using the smallest size of data that conveys the maximum information. Dimensionality reduction is not often performed in the context of space weather. A recent example is the use of Principal Component Analysis (PCA) for the nowcasting of solar energetic particles \citep{papaioannou2018}.

\begin{figure}[h]
 \centering
 \includegraphics[width=\columnwidth]{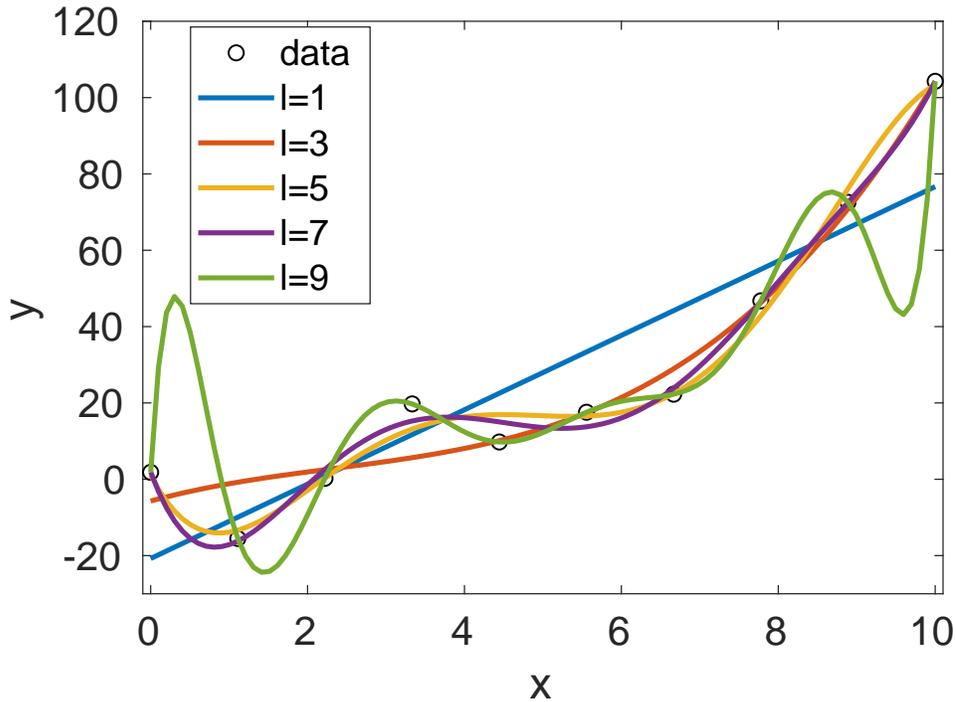}
 \caption{Example of overfitting with polynomial regression. By increasing the order of the polynomial $l$, the error with respect to the training data decreases (until for $l=9$ the data points are fitted exactly), but the model becomes less and less generalizable to unseen data. For reference, the data was generated as a cubic function of $x$ with small Gaussian noise.}
 \label{fig:polyfit}
  \end{figure}

{
\section{Machine Learning workflow}\label{sec:ML_work}
In this final Section before the review part of the paper, we summarize the different phases that constitute the workflow in applying machine learning to a space weather problem (and maybe more generally to any physics problem). This is not to be considered as a strict set of rules, but rather as a guideline for good practice. This workflow is inspired by the \emph{scikit-learn algorithm cheat sheet} (\url{https://scikit-learn.org/stable/tutorial/machine_learning_map/}).

\subsection{Problem formulation}
The importance of formulating the problem in a well-posed manner cannot be overstated. The relative easiness of using an off-the-shelf machine learning library poses the serious risk of trying to use machine learning for problems that are not well formulated, and therefore whose chances of success are slim.
It is not straightforward to define what a well-posed problem is. First, one has to define what is the objective of the study, and to address a number of questions related to the well-posedness of the problem:
\begin{itemize}
 \item Predict a quantity: Regression (see Sec. \ref{sec:regression})\\
 Is there any physical motivation that guides us into choosing the independent variables?\\
 Are time dependence and causality taken into account? Forecasting or Nowcasting?\\
 Do we have enough data so that the trained algorithm will be generalizable?\\
 Is the uniqueness of the input-output mapping physically justified?\\

 \item Predict a category 
 \begin{itemize}
  \item Labels are known: Supervised Classification (see Sec. \ref{sec:supervised_class})\\
  Are the labeled classes uniquely defined and disjoint?\\
  Do we expect to be controlling variables that uniquely define the boundary between classes?\\
  Is the data balanced between classes?\\  
  
  \item Labels are not known: Clustering (see Sec. \ref{sec:clustering})\\
  Is there a physical reason for the data to aggregate in clusters?\\
  Do we have a physical understanding of what is the optimal variables space where clustering becomes more evident?\\
  Do we expect to be able to physically interpret the results obtained by the clustering algorithm?\\
  Is the data representative of all the clusters we might be interested into?\\
 \end{itemize}
 
 \item Discover patterns or anomalies in the data: Dimensionality reduction (see Sec. \ref{sec:dim_red})\\
 Is there a physical motivation that can guide our expectation of the optimal dimensionality?\\
 Are there variables that are trivially redundant or strongly correlated?
 
\end{itemize}

\subsection{Data selection and pre-processing}
The quality of the data will largely affect the goodness of a machine learning algorithm. After all, machine learning constructs a non-trivial representation of the data, but it will not be able to find information that is not contained in the data in the first place. This is the step where a specific domain expertise and collaboration with the persons responsible of the data management (for instance, the PI of a satellite instrument) becomes very important. From an algorithmic point of view, data pre-processing involves so-called exploratory data analysis, that consists in collecting descriptive statistics (probability distribution, percentile, median, correlation coefficients, etc.) and low-dimensional visualization that is descriptive of the data (heat maps, scatter plots, box plots, etc.). In this step human intuition can still play a role in steering the machine learning workflow towards the most effective algorithm. \\}
It is also worth mentioning a whole field of research devoted to understand causal relationship between physical observables, that uses tools adopted from Information Theory. A whole review could be devoted to that topic, and here I will only uncover the tip of the iceberg. For a recent review, we refer the reader to \citet{johnson2018}.
In short, within the field of System Science, Information Theory can be used to address the question: what is the smallest possible (i.e. not redundant) set of variables that are required to understand a system? Using ideas based on the well-known Shannon entropy \citep{shannon1948}, one can define Mutual Information as the amount of information shared between two or more variables, one can look at cumulant-based cost as a measure of nonlinear dependence between variables, and finally infer their causal dependence by studying their transfer entropy.
For instance, \citet{wing2016} have studied the relationship between solar wind drivers and the enhancement of radiation belt electron flux, within a given time-lag. This approach, not only is able to rank the proposed drivers in terms of importance, but it also provides a maximum time horizon for predictions, above which the causal relationship between inputs and outputs becomes insignificant. This is extremely valuable in designing a forecasting model, because it informs the modeler on what inputs are physically relevant (hence avoiding to ingest rubbish in).
Other studies of space weather relevance are \citet{johnson2005, materassi2011, wing2018}.\\
{
Pre-processing also involves data cleaning and taking care of any data gaps one might encounter. Unfortunately, the way data gaps are handled (for instance gaps can be filled by interpolation, or data with gaps can be discarded) can affect the final outcome. Also, one has to think of how to deal with any outliers. Are outliers physically relevant (and maybe the extreme events we are interested in predicting) or just noise?
And finally, one might consider if it makes sense to augment the data to reduce imbalance or improve the signal-to-noise ratio (see also Sec. \ref{sec:new_trends}).

\subsection{Algorithm selection}
The choice of the most promising algorithm depends on a number of factors. Unfortunately, this is the area where the science overlaps with the art. One interesting consideration is that, in theory, there is no reason for one algorithm to outperform other algorithms: when interpreted as optimization problems, a local minima of a chosen cost function should be detected as a local minima by \emph{any} algorithm. However, in practice, the internal working of a given algorithm is related to a particular choice of the free parameters (hyper-parameters), and one cannot fully explore the hyper-parameter space.
Hence, algorithm selection often boils down to a trade-off between accuracy, training time, and complexity of the model. \\
Other considerations involve whether the model needs to be regularly re-trained (for instance with incoming new data like in the model of \citet{ling2010} discussed in Section \ref{sec:GEO}), how fast the model runs in prediction mode (after being trained), and whether it is scalable with respect to increasing the dataset size. For a more detailed discussion about where each machine learning algorithm stands in terms of accuracy, computational cost, scalability, we refer the reader to specialized textbooks.\\ However, there is one simple concept that is useful to introduce, that divides the algorithms in two camps: parametric vs non-parametric. Models that have a fixed number of parameters are called parametric, while models where the number of parameters grow with the amount of training data are called non-parametric . The former have the advantage of being faster to train and to be able to handle large dataset. The disadvantage is that they are less flexible and make strong assumptions about the data that might not be appropriate. On the other hand, non-parametric models make milder assumptions, but are often computationally intractable for large (either in size or in dimensions) datasets \citep{murphy2012}. Examples of parametric models include linear and polynomial regressions and neural networks. Non-parametric models include k-means and kernel methods such as Gaussian Processes, Support Vector Machines, and kernel density estimators.

\subsection{Overfitting and model selection}
After selecting a machine learning algorithm, the next step consists in training the model, that is to optimize its parameters. Yet there are a number of parameters, dubbed hyper-parameters that are free to choose (that is, their value is not a result of an optimization problem). Appropriately tuning the hyper-parameters can have a non-negligible impact on the accuracy and computational cost of training a model. Moreover, in parametric models the number of hyper-parameters is itself a degree of freedom (for instance, the number of neurons in a neural network). Model selection deals with the choice of hyper-parameters.\\}
It is also important to stress out the concept of overfitting, which is frequently invoked as a weakness of machine learning, but often inappropriately.
The idea can be easily understood by analyzing polynomial regression in one dimension. Let us assume to have 10 data points that we want to approximate by means of a polynomial function. Recalling our nomenclature in definition (\ref{def_y}), $f(x)=\sum_l a_lx^l$ (where $l$ is now an exponent and the index of the unknown vector of coefficients $\mathbf{a}$). In principle, one can always find the 9th order polynomial that fits exactly our 10 points, for which the model error $\varepsilon=0$, no matter how it is  defined. However, this would result in a highly-oscillatory function that will unlikely pass close to any new data point that we will observe in the future, and rapidly diverging outside the range of the initial data points (see Figure \ref{fig:polyfit}). 

This is a simple example of data overfitting, where the underlying function was made fit the noise {rather} than the signal, reducing the error $\varepsilon$ to zero, when calculated on the training set. On the other end of the spectrum in polynomial regression, one might equally be unhappy with using a simple linear function, as the one described in Section (\ref{sec:regression}), that might not be able to capture, for instance, a faster than linear increase in $x$. Eventually, the problem we face is a trade-off between the complexity of the model, that is its ability to capture higher-order nonlinear functions, and its ability to generalize to unseen data. This problem is common to any machine learning algorithm, where the complexity (number of hyper-parameters) can be chosen and fine-tuned by the user. For instance, in a neural network, a larger number of neurons and hidden layers determine its ability to approximate {more and more complex} functional forms. The risk is to convince ourselves to have devised a very accurate predictor that effectively is not able to predict anything else that what has been fed as training data.

\subsubsection{Training and validating}
Several strategies exist to circumvent this misuse of machine learning algorithms.
Unfortunately, they all come at the cost of not using the entire wealth of data at our disposal, and to sacrifice some of that. In practice, one divides the available data into three disjoint sets: training, validation and test.
The training set is used to effectively fine-tune the many unknown parameters that constitute the model. Algorithms are commonly trained iteratively by one of the many variants of a stochastic gradient descent method \citep{ruder2016}, that seeks to reduce the value of the cost function at each iteration by updating the unknown parameters that enters in the definition of the chosen cost function. Especially for not very large datasets, one can push such minimization to very low values of the cost function, which corresponds to an over fit on the training set. In order to avoid overfitting, the cost function is periodically evaluated (every few iterations) on the validation set. Because the algorithm does not use these data (validation) in the minimization of the cost function, this should not decrease unless the method has captured some generic features of the data that are not specific to the training set. In practice what happens is that both cost functions evaluated on the training and validation sets decrease (on average) for a certain number of iterations, until at some point the cost calculated on the validation set stops decreasing and starts increasing. That is a sign that the algorithm is starting to pick features that are distinctive of the training set and not generalizable to the validation set. In other words, it is starting to fit the noise, and the iterations should be stopped. At that point, further reducing the score on the validation set (for the same amount of model complexity) would probably require more information in terms of latent variables.

\subsubsection{Cross-validation}
Another procedure that is often used in machine learning is called cross-validation \citep{shao1993,schaffer1993}. In order to assure that a given model is not specific to an arbitrary choice of a training set and that its good performance is not just good luck, one can split the original training set into $k$ disjoint partitions and use $k-1$ of them as training set and the remaining one as validation set. By permuting the role of validation and training, one can train $k$ different models, whose performance should approximately be equal and whose average performance can be reported. 

\subsubsection{Testing and metrics}
Finally, the test set plays the role of `fresh', unseen data on which the performance metrics should be calculated and reported once the model has been fine-tuned and no further modifications will be done. A few subtle pitfalls can be encountered using and defining the three sets. For instance, in the past it was common to split a dataset randomly, while it is now understood that if temporal correlations exist between events (which always exist in the common case of time series of observations) a random split would result in an artifactual increase of performance metrics for the simple reason that the unseen data in the validation set is not truly unseen, if it is very similar to events that belong to the training set, because temporally close.
Another pitfall concerns the fine-tuning or the choice of a model a posteriori, that is after it has been evaluated on the test set. Let as assume that we have two competing models that have been trained and validated. Any further information that is gained by evaluating the models on the test set should not be used to further improve the models, or to assess which model performs better. \\
{Both the final performance and the cost function are represented in terms of metrics. It is a good practice to use different metrics for the two purposes. In this way one can assure that the model performs well with respect to a metric that it was not trained to minimize, hence showing robustness. 
We report a list of performance metrics and cost functions routinely used for regression and classification, both in the deterministic and probabilistic case, in Table (\ref{table:metrics}). A useful concept is that of \emph{skill score} where the performance of a model is compared with respect to a baseline model. Usually, the baseline is chosen as a zero-cost model, such as a persistence or a climatological model. 
 For extensive discussions about metric selection, the reader is referred to \citep{morley2018, bloomfield2012, bobra2015, liemohn2018b}}

\subsubsection{Bias-Variance decomposition}
{The mentioned trade-off between complexity and ability to generalize can be understood mathematically by decomposing the error in what is known as bias-variance decomposition. The bias represents the extent to which the average prediction over all data sets differs from the desired outcome. The variance measures the extent to which the solutions for individual data vary around their average or, in other words, how sensitive a model is to a particular choice of data set \citep{bishop2006}. Very flexible models (more complex, many hyper-parameters) have low bias and high variance and more rigid models (less complex, few hyper-parameters) have high bias and low variance.
Many criteria exist that help select a model, by somehow penalizing complexity (for instance limiting the number of free parameters), such as the Bayesian Information Criterion (BIC) \citep{schwarz1978}, the Akaike Information Criterion (AIC) \citep{akaike1998}, and the Minimum Description Length \citep{grunwald2007}. This is a wide topic and we refer the reader to more specialized literature.\\
} 
 \begin{longtable}{p{5cm}|  p{4cm} | p{5cm}}
  \caption[]{Performance metrics for binary classification and regression, both for deterministic and probabilistic forecast.}\label{table:metrics}\\
  Performance metric & Definition & Comments\\
 \hline
 \multicolumn{3}{c}{Binary Classification - Deterministic} \\
  \hline
 Sensitivity, hit-rate, recall, true positive rate & $TPR = \frac{TP}{P}$ & The ability to find all positive events. Vertical axis in the ROC curve (perfect TPR = 1)\\  
 \hline
 Specificity, selectivity, true negative rate & $TNR = \frac{TN}{N}$ & The ability to find all negative events.\\
 \hline
 False positive rate& $FPR = \frac{FP}{N}=1-TNR$ & Probability of false alarm. Horizontal axis in Receiver Operating Characteristic (ROC) curve (perfect FPR=0).\\
 \hline
 Precision, positive predicted value & $PPV = \frac{TP}{TP+FP}$ & The ability not to label as positive a negative event (perfect PPV=1).\\
 \hline
 Accuracy & $ACC =\frac{TP+TN}{P+N}$ & Ratio of the number of correct predictions. Not appropriate for large imbalanced dataset (e.g. $N>>P$).\\
 \hline
 F1 score & $F1 = \frac{2 PPV\cdot TPR}{PPV+TPR}$ & Harmonic mean of positive predicted value (precision) and true positive rate (sensitivity), combining the ability of finding all positive events and to not mis-classify negatives.\\
 \hline
 Heidke Skill Score (1) & $HSS_1 = \frac{TP+TN-N}{P} = TPR\left(2-\frac{1}{PPV} \right)$ & It ranges between $-\infty$ and 1. Perfect $HSS_1=1$. A model that always predicts false can be used as a baseline, having $HSS_1=0$.\\
 \hline
 Heidke Skill Score (2) & $HSS_2 = \frac{2 (TP\cdot TN)-(FN\cdot FP)}{P(FN+TN)+ N(TP+FP)}$ & It ranges between -1 and 1. Skill score compared to a random forecast.\\
 \hline
 True Skill Score & $TSS = TPR -FPR = \frac{TP}{TP+FN} - \frac{FP}{FP+TN}$ & Difference between true and false positive rates. Maximum distance of ROC curve from diagonal line. Ranges between -1 and 1. It is unbiased with respect to class-imbalance.\\
 \hline
 \multicolumn{3}{c}{Binary Classification - Probabilistic} \\
 \hline
 Brier score & $BS = \frac{1}{N}\sum_{i=1}^N(f_i-o_i)^2$ & $N$ is the forecast sample size, $f_i$ is the probability associated to the event $i$ to occur, $o_i$ is the outcome of event $i$ (1-true or 0-false). Ranges between 0 and 1. Negatively oriented (i.e. perfect for BS=0).\\
  \hline
  Ignorance score & $IGN = \frac{1}{N}\sum (o_i-1 )\log(1-f_i) -o_i\log(f_i)$ & Definitions as above, except IGN ranges between 0 and $\infty$.\\
  \hline
 \multicolumn{3}{c}{Continuous Variable (Regression) - Deterministic} \\
 \hline
 Mean Square Error & $MSE = \frac{1}{N}\sum_{i=1}^N(\hat{y}_i - y_i)^2$ & $N$ is the size of the sample, $\hat{y}_i$ is the $i$-th prediction (scalar real value) and $y_i$ is the corresponding observation. MSE penalizes larger errors (sensitive to outliers).\\
 Root Mean Square Error & $RMSE=\sqrt{MSE}$ & It has the same units as $y$\\
 Normalized Mean Square Error & $NRMSE=\frac{RMSE}{\overline{y}}$ & $\overline{y}$ is either defined as the mean of $y$ or its range $y_{max}-y_{min}$\\
 \hline
 Mean Absolute Error & $MAE = \frac{1}{N}\sum_{i=1}^N|\hat{y}_i - y_i|$ & MAE penalizes all errors equally: it is less sensitive to outliers than MSE.\\
 \hline
 Average Relative Error & $ARE = \frac{1}{N}\sum_{i=1}^N\frac{|\hat{y}_i - y_i|}{|y_i|}$ & \\
 \hline
 Correlation coefficient & $cc$ or $R=\frac{\sum_{i=1}^N (\hat{y}_i - \mu_{\hat{y}})({y}_i - \mu_{{y}})}{\sqrt{\sum_{i=1}^N (\hat{y}_i - \mu_{\hat{y}})^2 } \sqrt{\sum_{i=1}^N (y_i - \mu_{y})^2 }}$ & $\mu_{\hat{y}}$ and $\mu_y$ are respectively the mean values of the predictions $\hat{y}$ and of the observations $y$. R ranges between $-1$ (perfect anti-correlation) to $1$ (perfect correlation)\\
 \hline 
 Prediction Efficiency & $PE = 1-\frac{\sum_{i=1}^N(\hat{y}_i - y_i)^2}{\sum_{i=1}^N(y_i - \mu_y)^2}$ & Perfect prediction for $PE=1$\\ 
 \hline
 Median Symmetric Accuracy & $\zeta = 100(\exp(M (|\log{Q_i}|))-1)$ &  $Q_i = \hat{y}_i/y_i$ and $M$ stands for Median. See \citep{morley2018b}\\
 
 \hline
  \multicolumn{3}{c}{Continuous Variable (Regression) - Probabilistic} \\
  \hline
  Continuous Rank Probability Score & $CRPS = \frac{1}{N}\sum_i \int_{-\infty}^\infty (\hat{F}_i(z) - H(z-y_i))^2 dz$ & $N$ is the size of the sample, $\hat{F}_i(y)$ is the $i$-th forecast probability cumulative distribution function (CDF), and $H$ is the Heaviside function. CRPS collapses to MAE for deterministic predictions, and it has an analytical expression for Gaussian forecast \citep{gneiting2005}.\\
  \hline
  Ignorance score & $I(p,y) =\frac{1}{N}\sum_i -\log(p_i(y_i))$ & $p_i(y_i)$ is the probability density function associated to the $i$-th forecast, calculated for the observed value $y_i$ \\
  \hline
  \multicolumn{3}{p{\textwidth}}{In binary classification (deterministic) $P$ and $N$ are the total number of positives and negatives, respectively and $TP$, $TN$, $FP$, $FN$ denote true-positive/negative and false-positive/negative. For probabilistic binary classification, $f$ is the forecasted probability and $o$ is the real outcome (1-true or 0-false). For deterministic regression, $y$ is the observed real-valued outcome and $\hat{y}$ is the corresponding prediction.}\\
  \hline
 \end{longtable}

\section{Review of machine learning in Space Weather}\label{sec_review}
\begin{figure}[h]
 \centering
 \includegraphics[width=\columnwidth]{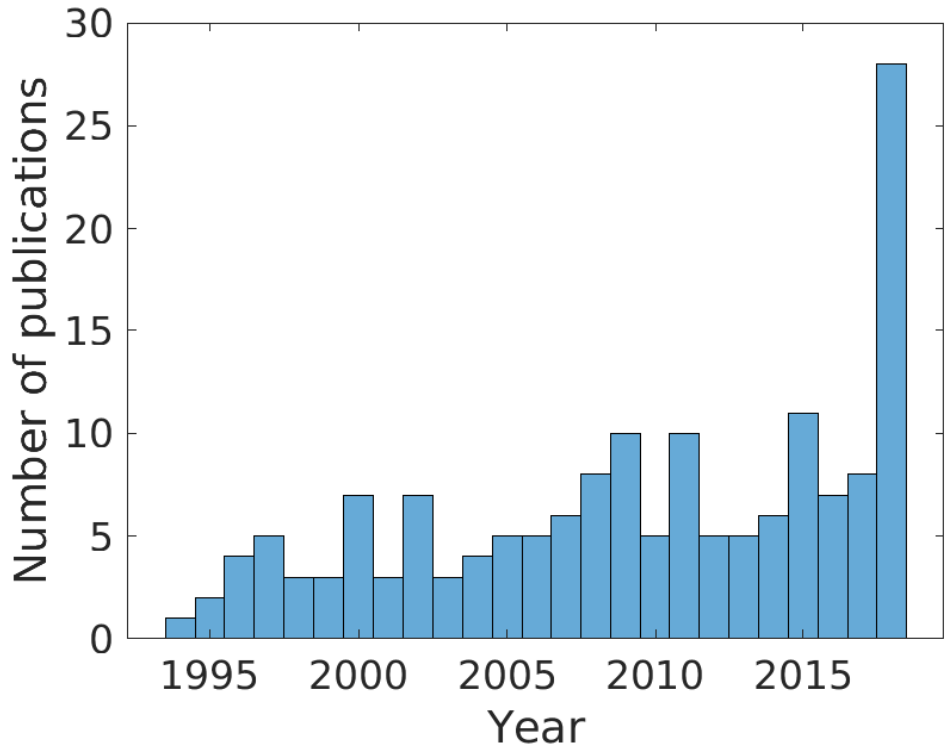}
 \caption{Number of publications between 1993 and 2018 in the area of machine learning applied to space weather cited in this review.}
 \label{fig:hist}
  \end{figure}

In this Section we review some of the literature concerning the use of machine learning in space weather. We focus our attention on three applications that seem to have received most scrutiny: the forecast of geomagnetic indices, relativistic electrons at geosynchronous orbits, and solar eruptions (flares and coronal mass ejections). This review has no pretension of completeness, and as all reviews is not free from a personal bias. However, the intention is to give an idea of the wide breadth of techniques covered over the years, more than to offer detailed comments on specific works. Also, even if we report performance metrics, it has to be kept in mind that an apple to apple comparison is often not possible, because different techniques have been tested on different datasets. Finally,  Figure \ref{fig:hist} emphasizes the timeliness of this review, by showing the distribution of publication years of the works cited in this paper (only the papers presenting a machine learning technique for space weather). The explosion of interest that has occurred in 2018 (the last bar to the right) is quite remarkable. Time will tell if that was just noise in the data.

\subsection{Geomagnetic indices}

A geomagnetic index is a simple measure of geomagnetic activity that attempts to condense a rich set of information about the status of the magnetosphere in a single number. Many such indices exist: historically $Kp$ and $D_{st}$ are probably the most widely used, but many more have been proposed (AE, AL, AU, ap, am, IHV, Ap, Cp, C9, SYMH, ASYH) \citep{rostoker1972, menvielle2011}. Each index is meant to capture a different aspect of geomagnetic activity, such as local geographical dependency. An interesting attempt to construct a single composite index that would uniquely define the geomagnetic state has been recently proposed in \citet{borovsky2018}.\\

The prediction of a geomagnetic index has always been a very attractive area for machine learning applications, because of its straightforward implementation, the well-posed definition of indices, the availability of large historical dataset, and the restricted range of possible outcomes.
$D_{st}$ and $Kp$ are the ones that have received most attention, with the first models proposed in \citet{lundstedt1994, gleisner1996, wu1997}.

\subsubsection{Forecasting $Kp$}

The use of a neural network to forecast $Kp$ either one or multiple hours in advance has been proposed in \citet{costello1998, boberg2000, gholipour2004, wing2005, valach2006, bala2009, uwamahoro2014, tan2018, wintoft2017}, among others. 
Real-time forecasts based on some of these models are running at RWC, Sweden (\url{http://www.lund.irf.se/forecast/kp/}), Rice Space Institute, USA (\url{http://mms.rice.edu/mms/forecast.php}), INPE, Brazil (\url{http://www2.inpe.br/climaespacial/portal/swd-forecast/}), the Space Environment Prediction Center, China (\url{http://eng.sepc.ac.cn/Kp3HPred.php}).\\

The US Space Weather Prediction Center (SWPC/NOAA) has provided real-time one and four hours ahead forecast based on the \citet{wing2005} model from 2010 until 2018, when the Wing $Kp$ was replaced by the physics-based Geospace model developed at the University of Michigan \citep{toth2012}. 
The Wing $Kp$ model used solar wind parameters at L1 ($|V_x|$, density, IMF $|B|$, $B_z$) and the current value of $Kp$ to predict the future $Kp$ approximately one hour ahead (a modified model predicted 4 hours ahead). By comparing with the competing models of the time (i.e, the models by \citet{costello1998} and \citet{boberg2000} and the NARMAX model \citep{ayala2016, boynton2018}), \citet{wing2005} reported an higher performance, attributed to a larger training set, and the inclusion of nowcast $Kp$, which is highly correlated with its future values, and a correlation coefficient $R=0.92$. However, the authors noticed that this metric, by itself, does not indicate how well a model performs. 

Because $Kp$ is a discrete index, one can look at metrics designed for discrete events that take into account the number of true/false positive/negative. One such metrics is the True Skill Score (see Table \ref{table:metrics}) that was considered in \citet{wing2005}, where they reported a $TSS\sim 0.8$ for the range $2\leq Kp \leq 8$. They considered both feed-forward and recurrent neural networks, with one hidden layer and the number of hidden neurons ranging between 4 and 20. The dataset covered the period 1975-2001, that was randomly split into training and a test sets of equal size. It is now realized that a random split is not the best procedure, since the test set (on which the final metrics are reported) gets contaminated by the training set. In other words, the two sets are not completely independent and the reported performance is higher than if it was calculated on out of samples (unseen) data. 

A parallel, independent effort has been carried out by the Rice Space Institute, which provides real-time 1h and 3 h forecast \citep{bala2009, bala2012}. These predictions are also based on neural networks, with the interesting characteristic of using coupling functions (and their history) as inputs. The original work used only the Boyle index (BI), which empirically approximates the polar cap potential as a function of solar wind speed, magnitude of interplanetary magnetic field, and clock-angle \citep{boyle1997}. Improved models also included dynamics pressure. Comparing the use of BI,  Newell, and Borovsky coupling functions \citep{newell2007, borovsky2008} resulted in very similar forecasting performance, with Newell having slightly better metrics (correlation coefficient, root-mean-square-error, and average relative error). This result seems to be in line with the general idea that neural networks are universal approximators and, given enough expressive powers (in terms of number of hidden layers and number of neurons), they should be able to transform the inputs into any complex nonlinear function. Hence, the question arises of how beneficial it is to feed the network with a given coupling function, rather than the individual physical quantities that enters in such function, and that might just depend on how deep the network is, or on the numbers of neurons for single hidden layers networks.

\citet{ji2013} proposed to improve past work based on neural networks, by also including all three components of interplanetary magnetic field and the $y$  component of electric field. They reported higher performance with respect to the models by \citet{costello1998,bala2012,wing2005}, however the comparison was not carried out with equal network architecture or same training and test dataset.

The model of \citet{boberg2000} was recently improved in \citet{wintoft2017}. The main innovations with respect to the original work are the inclusion of local time and day of the year as inputs and the use of an ensemble of networks. Also, the model was designed not to forecast $Kp$ with a pre-fixed lead time (i.e. one hour ahead), but by using a variable propagation lead time that depends on the solar wind velocity. As a results, the lead times range between 20 and 90 mins. Although this might seem more accurate, it brings in the additional difficulty of accurately estimating the solar wind propagation time, and to quantify the uncertainties associated with such estimate. The reported performance was $RMSE\sim 0.7$ and correlation coefficient $cc\sim 0.9$. 

Some very interesting elements of novelty in the use of NN to forecast $Kp$ have been presented in \citet{tan2018}. Following the current trend of 'going deep', and leveraging of recent advances in neural networks, they proposed to use a Long Short-Term Memory network (LSTM, \citet{hochreiter1997, gers1999}). This is a special type of recurrent network, and its main characteristic is the ability of retaining information from the past, being able to automatically choose the optimal time lag, that is how long back in time the information is still relevant. LSTM has been successfully employed in many fields of time-series forecasting \citep{goodfellow2016}. They also discuss the well-known problem of data-imbalance, meaning that the distribution of $Kp$ is highly skewed, with a typical ratio of storm to non storm close to 1:30. 
The main feature that differentiate this work from all the previous papers, is the idea of first casting the problem into a classification task, namely to predict whether the next 3 hours fall in the storm ($Kp \geq 5$) or quite ($Kp <  5$) condition. They then train two separate regression sub-models for each case. Hence the prediction pipeline is made of a classification step, that decides which sub-model for regression is called. Obviously each sub-model is trained only on the relevant dataset. This can be seen as a special case of ensemble modeling (with only two members), where the final outcome is not an average of all ensemble predictions, but rather a \emph{winner takes all} model. The apparent downside is that any mis-classification in the pipeline will likely result in a bad performance of the regression sub-models. The authors studied the correlation between 11 candidate input parameters and eventually (probably also due to the heavy load of the LSTM training) chose only three inputs: proton density, $Kp$ and the Boyle index $BI$. The final metrics are not overwhelmingly superior to previous works: $RMSE = 0.64$ and $cc = 0.81$.

A methodology based on Nonlinear Autoregressive with Exogenous inputs (NARX) was presented in \citet{ayala2016}. 
This family of models is not very dissimilar from NN, in that the expected output is modeled as a superposition of nonlinear functions of the inputs. In NARX, such nonlinear functions are taken as a large combination of monomials or polynomials of inputs, including error terms. In principle, one could retain a large number of terms, however in practice the vast majority of monomials will have no influence on the output. One of the objectives of a NARX model is to identify a parsimonious combination of inputs. An algorithm to identify the most important terms is the so-called FROLS (Forward Regression Orthogonal Least Square) algorithm \citep{billings1989, billings2013}, that is used in combination with the Error Reduction Ratio (ERR) index to measure the significance of each candidate model term. In \citet{ayala2016} six terms were eventually identified as input drivers: past $Kp$ values, solar wind speed, southward interplanetary magnetic field, the product of the two, solar wind pressure, and its square root. Several models were proposed and tested, for a different range of prediction time lag, using 6 months of data from the year 2000 for training and 6 months for testing. However, only one model provided a true (3 hours ahead) forecast, that is not using future values of some input. That models resulted in the following performance: $RMSE\sim 0.8$, $cc\sim 0.86$, $PE\sim0.73$. In particular, the authors noted a consistent bias in under-predicting events with $Kp\geq 6$.

Finally, the recent work by \citet{wang2015} stands out in providing a probabilistic forecast (rather than a single-point prediction), by constructing conditional probabilities over almost 40,000 3-hourly events in the period August 2001 - April 2015. The authors have tested more than 1,200 models by considering different combination of three conditional parameters, among a possible choice of 23 inputs. They cast the problem as a classification task that forecasts the category of $Kp$ rather than its value (the 28 discrete values are grouped into four categories: quiet, active, minor storm, strong storm). The performance of the models is appropriately measured in terms of Rank Probability Score (RPS), Discrimination Distance (DISC), and relative operating characteristic area (ROCA). The best performing model yields an RPS value equal to 0.05, which is about half of what results by using a classical climatological model.
Hence, this model can provide a simple and effective baseline to test future probabilistic predictions.

\subsubsection{Forecasting $D_{st}$}

The $D_{st}$ index is based on four low-latitude stations and it measures the deviation of the horizontal component of the Earth's magnetic field from its long term average. It is  a proxy for the axi-symmetric magnetic signature of magnetospheric ring currents \citep{sugiura1963}. It is an hourly-based index, measured in nT, and it can be considered a continuous value index, even though it is expressed as an integer, with minimal increments of 1 nT.

As already mentioned, the forecasting of $D_{st}$ has been the subject of intensive investigation using machine learning techniques.
\citet{wu1996, wu1997} presented one of the first application of artificial neural networks for one to eight hours ahead forecasts. They have proposed the use of a two-layer network with feedback connection (Elman architecture, \citep{elman1990}) which was designed to capture time-dependent dynamics. They tested a combination of solar wind parameters inputs, including speed, density, total magnetic field and its southward component, and products of them. They used a dataset covering years 1963-1992. The best performing network yielded a correlation coefficient $cc\sim 0.9$, root-mean-square-error $RMSE\sim 15$, and prediction efficiency $PE\sim 0.8$, for one hour ahead. The general conclusion for predictions more than one hour ahead was that the initial phase of a storm was not accurately predicted, while the main phase could be predicted relatively well up to two hours in advance.
This model was further improved and made operational (for one hour ahead) in \citet{lundstedt2002}. A remarkable features is that the trained network is extremely compact (especially compared to today standards), with only 4 hidden layer neurons. The values of weights and bias were given in the paper, and relative scripts are available on \url{http://lund.irf.se/rwc/dst/models/}. 

\citet{kugblenu1999} have improved the prediction performance of one hour ahead forecast, by including the 3-hours time history of $D_{st}$, and achieving a performance efficiency $PE$ as high as 0.9. However, they trained and tested their network exclusively on storm times (20 storms for testing and 3 storms only for testing).

\citet{pallocchia2006} made the interesting argument that in-situ solar wind plasma instruments tend to fail more often than magnetometers, because they can saturate for several hours due to large emission of particles and radiation. This can be problematic for operational forecasting based on solar wind density and velocity. For this reason, they proposed an algorithm based exclusively on IMF data and the use of an Elman network, dubbed EDDA (Empirical Dst Data Algorithm). Somewhat surprisingly, they reported a performance comparable to the Lund network (with the caveat that training and test sets were different, 58,000 hourly averages used for EDDA and 40,000 for Lund). An interesting test was shown on the 2003 Halloween storm, when the ACE/SWEPAM instrument malfunctioned for several hours, transmitting incorrect values of density and bulk flow speed, while the ACE/MAG magnetometer continued to produce reliable data. In this situation the Lund operational forecast becomes unreliable, while EDDA still produces valid predictions. 

\citet{voros2002} have made the interesting suggestion of using the information about the scaling characteristics of magnetic fluctuations as an additional input to a neural network. They have implemented this by computing the so-called H\"{o}lder exponent of past geomagnetic time series and shown that it significantly improved the prediction accuracy. They also expanded the standard set of inputs by including time derivatives of magnetic field intensity, solar wind speed and density, and performed a dimensionality reduction of inputs by using Principal Component Analysis (PCA), effectively reducing the number of inputs to two.
A related idea has been more recently proposed in \citet{alberti2017}, where the timescales associated with solar wind-magnetospheric coupling have been investigated through an Empirical Mode Decomposition (EMD), with the suggestion that information relevant at different timescales (i.e. above or below 200 mins) can directly be used for geomagnetic forecasting.

\citet{lethy2018} have presented an extensive study on the geoeffectiveness of different combinations of solar wind parameters, on the effect of different training set periods, and of different prediction horizon and time delays, using a single layer neural network. They have presented results covering 1 to 12 hours ahead predictions, and reporting  $RMSE\sim 12$  and $cc\sim 0.9$ for 12 hours ahead forecast (tested on a few storms in the period 2016-2017). The authors remark that their method has slightly lower accuracy than other methods for short-time prediction, but that it stands out in medium-term (12 hours) prediction. 

The standard method of training a neural network is by using a so-called back-propagation algorithm, where the iterative update of weights and biases are calculated by using information on the gradient (that is calculated analytically in a neural network) and the repeated application of the chain rule for derivatives \citep{care2018}. Other methods exist, based on global optimization techniques including simulated annealing, genetic algorithms, and particle swarm. The latter method has been proposed in \citet{lazzus2017}, for training a feed-forward NN with a single hidden layer containing 40 neurons. The particle-swarm technique has the advantage of being less sensitive to the weights initial conditions, and less prone to being `stuck' in local minima during training. In this work, the authors used inputs composed of the time history of $D_{st}$ only and a remarkably large dataset for training, validation and test sets (1990-2016). Six different models for forecasting $D_{st}$ one to six hours ahead were trained. Predictions up to three hours ahead yielded relatively high accuracy when specifically tested on 352 geomagnetic storms (the metrics, however, were calculated on the whole dataset including the training set): they reported a $RMSE\sim 10.9$ for one hour ahead, and $RMSE\sim 25$ for three hours ahead predictions.

Yet a different method to train a neural network, based on a Genetic Algorithm (GA) has been presented in \citet{vega2018}, where one to six hours ahead predictions were developed using a single hidden layer NN. The results were very good for one hour ahead, but degraded strongly for 6 hours ahead ($RMSE\sim 14$). A GA approach was also proposed in \citet{semeniv2015}.

The majority of the machine learning approaches to forecasting geomagnetic indices use neural networks. However, other machine learning techniques have been proposed. \citet{lu2016} have compared the use of Support Vector Machine (SVM, \citet{vapnik2013}) with neural networks. They have identified 13 solar wind input parameters, trained and tested their models on 80 geomagnetic storms (1995-2014). K-fold cross validation was used, meaning that 1/5 of the dataset (i.e., 16 storms) was left out for testing, repeating the experiment five times with different training sets, and finally averaging the results. Their best model achieved a correlation coefficient $cc\sim 0.95$.

\citet{choi2012} used the value of $D_{st}$ to distinguish between geoeffective ($D_{st}<-50$) and non-geoeffective Coronal Mass Ejections (CME) and used a Support Vector Machine to forecast that feature. The input parameters for the SVM classification were the speed and angular width of CME obtained from SOHO/LASCO and the associated X-ray flare class. 106 CMEs in the period 2003-2010 were used for prediction, yielding an accuracy of 66\%.

\citet{wei2007} used an expansion in radial basis function (RBF) to model $D_{st}$ as function of the time history of solar wind dynamic pressure and the product of velocity and magnetic field amplitude. The RBF kernel was chosen as a multi scale squared exponential. A total of 10 inputs and 15 regressors were selected. The model presented a good performance, even though it was tested on a very limited portion of data (156 hours only).

A NARMAX approach has been proposed in \citet{boaghe2001} and \citet{boynton2011}. By employing the Error Reduction Ratio technique, they have inferred that the best coupling function between solar wind parameters and $D_{st}$ is $p^{1/2}V^{4/3}B_T\sin^6(\theta/2)$ and derived an expression to forecast one hour ahead $D_{st}$ as function of the past values of $D_{st}$ and of the history of the coupling function. The analytical expression is explicitly given in \citet{boynton2011}. Finally, the model was tested for 10 years of data (1998-2008) yielding a correlation coefficient $cc\sim 0.97$.

A NARX methodology has been compared to the use of Support Vector Regression (SVR) in \citet{drezet2002}, by using the seven-hours time history of $D_{st}$ and $VB_z$ only. The SVR method differs from other black-box approaches in the way it enforces parsimony (model complexity), by selecting a low-dimensional basis.

\citet{parnowski2008,parnowski2009} have used a simple linear regression approach, that yielded a prediction efficiency as high as $PE\sim 0.975$ for one hour ahead forecast and $PE\sim 0.9$ for three hours ahead. They used a statistical method based on the Fisher statistical parameter to calculate the significance of candidate regressors \citep{fisher1992}. The final total number of regressors was in the range 150-200. Aside from parameters whose geoeffectiveness is well understood (and used in previous model), one interesting result concerned the longitudinal flow angle, that was find to have a statistical significance larger than 99\%. 

\citet{sharifie2006} have proposed a Locally Linear Neurofuzzy (LLNF) model based on a Gaussian radial basis function for one to four hours ahead predictions. The model was trained using 10 years of data (1990-1999) and tested for a 6 months period, yielding $cc\sim 0.87$ and $RMSE\sim 12$ for 4 hours ahead.

Other methods include Relevance vector machine \citep{andriyas2015} and  Bayesian Neural Networks \citep{andrejkova2003}.

All the approaches discussed above fall in the category of supervised learning where a nonlinear mapping is sought between a set of inputs and the predicted $D_{st}$ output. An approach based on unsupervised learning has instead been proposed by \citet{tian2005}, based on the methodology of self-organizing maps (SOM) networks \citep{kohonen1997}. A SOM is a neural network where no `ground truth' output is provided and the objective of the network is to cluster similar events in a two-dimensional map, where the distance between neurons signifies a similarity between inputs. \citet{tian2005} have classified $E_y\sim VB_z$ into 400 categories, using a dataset covering the period 1966 - 2000. A total of 21 categories (neurons) have then been identified as indicators of geomagnetic storms, out of which 6 were connected to large storms (defined as $D_{st}\leq -180$ nT). Even though, this approach does not provide a predicted value for $D_{st}$ (i.e. it is a classification task, rather than a regression), it is still interesting to evaluate its performance in terms of predicting the occurrence of a storm. The authors identified 14 categories that provide a 90\% probability of intense storm, and the 6 categories associated with strong storms have a missing prediction rate of about 10\%. These are promising results, with the only drawback that the authors did not separately evaluate	 the performance on training and test sets (based on the argument that the training is unsupervised). Hence it would be interested to compute the prediction rate of the trained network on unseen data.

We finally turn our attention to probabilistic forecast. The overwhelmingly majority of methods provide a single-point estimate, with no indication of probabilities or uncertainties associated to the forecast.
However, the quantification of uncertainties and the understanding of how they propagate from data to models and between interconnected models is becoming a predominant theme in space weather, recently highlighted in \citet{knipp2018}. In fact, the operational and decision-making aspect of space weather depends largely on the uncertainty of a forecast, and on the reliability of such uncertainty.

\citet{chen1997} have introduced a Bayesian method to predict the geoeffectiveness of solar wind structures (defined as geoeffective when they result in $D_{st}<-80$), that has been subsequently tested for real-time WIND/IMF data covering the period 1996-2010 in \citet{chen2012}. Even though, strictly speaking, this is not a machine learning approach, it is still worth commenting, being one of the few real-time probabilistic predictions of $D_{st}$. 
In fact, although a large emphasis is given in the original paper on the physical features and recognition of magnetically organized structures, the method essentially employs a statistical analysis.
The original method considers the components of the magnetic field and the clock angle as sufficient features to obtain a large accuracy rate for moderate to large storms, while the inclusion of solar wind speed and density slightly improves the classification of weak storms. The method is a straightforward implementation of Bayes theorem using probability distribution functions constructed from the OMNI database covering the period 1973-1981. The output of the prediction is the estimated duration of an event and its associated probability to be geoeffective. A contingency table presented in \citet{chen2012} (where a probability is translated into a binary classification using 50\% as a threshold) shows an accuracy rate of 81\% (on a total of 37 storms). 

A more sophisticated probabilistic method, based on Gaussian Processes (GP) has been proposed in \citet{chandorkar2017} and \citet{chandorkar2018}. Gaussian process regression is a Bayesian method that is very appealing for its straightforward implementation and non-parametric nature. One assumes a certain covariance structure (kernel) between data points (that is between all pairs of training and test points) and predictions are made by calculating Gaussian probabilities conditioned on the training set. By working with Gaussian distributions the mathematical implementation is analytically tractable, and it boils down to simple linear algebra. The output is a prediction in terms of a mean and standard deviation of a Gaussian distribution. \citet{chandorkar2017} have tested a GP autoregressive model (using past history of $D_{st}$, solar wind speed and $B_z$ as regressors) on the set of 63 storms proposed in \citet{ji2012} for the period 1995-2006. They reported a $RMSE\sim 12$ and $cc\sim 0.97$ for one hour ahead prediction. 

A clear advantage with respect to parametric models, such as neural networks, is that the number of adjustable parameters (hyper-parameters, see Sec. \ref{sec:ML_work}) in a Gaussian Process is typically very small. On the other hand, a major drawback is the non-optimal scalability with the size of the dataset. To overcome the computational bottlenecks, sparse (approximate) Gaussian Process has been proposed and it has become a standard procedure in the Machine Learning literature (see, e.g., \citet{rasmussen2004}).\\
An interesting approach that combines the power and scalability of neural networks with the probabilistic interpretation of Gaussian Process has recently been presented in \citet{gruet2018}. In this work, an LSTM neural network is trained to provide up to 6 hours ahead prediction of $D_{st}$ using solar wind parameters and the magnetic field amplitude recorded by a GPS satellite. The neural network prediction is then used as a mean function for a GP regression problem, with the final outcome being a Gaussian probabilistic prediction. The model yields a $RMSE\sim 10$ and $cc\sim 0.9$ for six hours ahead predictions, with relevant information on the uncertainty of the predictions, even when tested for storm events.

\subsection{Recapitulation - Geomagnetic indices}\label{sec:recap_GEO}
It is evident that geomagnetic index prediction has served as a testbed for a plethora of machine learning techniques for the last 20 years. 
This short review is necessarily incomplete (for more related or similar works, see \citet{takalo1997, pallocchia2008, gavrishchaka2001, gavrishchaka2001b, gleisner1997, hernandez1993, revallo2014, dolenko2014, revallo2015, stepanova2008, stepanova2000, barkhatov2000, watanabe2002, watanabe2003,  srivastava2005, mirmomeni2006}). The reader might feel overwhelmed by the quantity and the diversity of published work.
Yet, it is not easy to formulate a clear answer to the question: how well are machine learning techniques doing in predicting geomagnetic indices? There are at least two main reasons: the first is that the body of literature has grown in an inorganic way, meaning that new works have not always built on previous results and experience and often new papers propose novel methods that are not straightforward to compare to early works.\\ 
The second reason is that the degree of freedom for any machine learning technique is quite large, in terms of the regressors to use and how long of a time history is appropriate, time horizon (how many hours ahead to predict), how to deal with data gaps, the time periods used for training, validation, and test, the cross-validation strategy, the metrics chosen to assess accuracy and reliability, and the complexity of a model (e.g., number of layers and neurons in a NN, hyper-parameters in kernel based methods).
The issue of the most appropriate choice of inputs is probably the topic that the most skeptics in the community use to criticize a machine learning approach. The argument is that by letting an algorithm choose what parameters are the most informative, with no regard for the physics behind it, one can risk to associate causal information to parameters that are actually not physically relevant and to develop a model that cannot distinguish very well the signal from the noise, or in other words that is not very able to generalize to unseen data (the proverbial `rubbish in - rubbish out'). In fact, the indisputable advantage of a physics-based model is that it will return a sensible result for any set of (sensible) inputs, and not only for a subset of seen data, as long as the assumptions and limitations of the model are valid.\\
Looking back at the evolution of machine learning models for geomagnetic indices, one can certainly notice that the early models were very cautious on choosing inputs and many papers provide physical argument to justify their choice. Also, there was a certain tendency (often not explicitly spelled out) to design parsimonious models, that is to have a trade-off between the complexity of the model and its accuracy. One reason is the notorious problem of over-fitting, again related to the lack of generality, but something to keep in mind to properly put in perspective models as old as five or ten years is that training a complex model was expensive. Nowadays, the advances in GPU computation and the availability of machine learning libraries that exploit GPUs with no effort for the user, have clearly moved the field into trying more complex models, the archetype of which are deep neural networks. 
The easiness of using open-access software for training a large (not necessarily deep) neural network is a double-edged sword. On one hand, it will allow us to explore increasingly complex models, in terms of number of inputs and nonlinear relationship among them, that were simply out of reach a decade ago. On the other hand, the `rubbish in-rubbish out' paradigm will always lurk in the indecipherable complexity of a model, even though to be completely fair I have not encountered, in preparing this review, a single work that uses a given input without providing even a vague physical motivation, simply because it seems to work!

\subsection*{What has not been done yet?}
The importance of being able of accurately predict geomagnetic indices several hours in advance is twofold. First, by incorporating some information of the Earth-magnetosphere system, geomagnetic indices give a warning on upcoming geomagnetic storms; second, they are often used to parameterize physical quantities in computational models. For instance diffusion coefficients in radiation belt quasi-linear models are often parameterized in terms of $Kp$ \citep[see, e.g,][]{tu2013}. Hence, the alleged superiority of physics-based models is severely weakened by their dependence on parameters empirical determined. \\
Most, if not all, previous works have focused on short- or medium-time prediction from solar wind drivers, often incorporating knowledge of the past state of the geomagnetic field, by using the same or other indices as input, or by using low or medium Earth orbit satellites \citep{gruet2018}. For physical reasons, these predictions cannot be made for horizon times longer than about 12 hours. 
In the future, we will see more attempts at forecasting indices directly from solar inputs that allow a prediction horizon of the order of days. For instance, \citep{valach2014} have presented a neural network for forecasting the C9 index based on geometrical properties of Coronal Mass Ejections (CME), such as position angle, width and linear velocity, and of observed X-ray flares, but still without using images.\\
The direct use of solar images and magnetograms will present a major challenge that will certainly be tackled in the near future, both in terms of data handling (with several Gbs of data at our disposal from SOHO and SDO) and in terms of the most optimal design of an accurate machine learning method. A deep convolutional neural network seems to be the most obvious choice (at least as a first attempt), given its well-documented ability of detecting features in images. However, there are many aspects that we do not currently know: do solar images contain enough information for predicting geomagnetic states? Would a one-step approach (from Sun to Earth) be feasible or should we envision a multi-step (Sun to L1, to magnetosphere, to Earth) similarly to what is done in modular physics-based simulations? Is the events imbalance (meaning a large abundance of quite time compared to a very few instances of storms, especially large storms) an insurmountable obstacle to the success of machine learning techniques, or we will be able to overcome it by augmenting data either empirically or through simulations? \\
I believe that the answer to most of these question will be established within the next decade.
And finally, how to incorporate physics knowledge into a machine learning algorithm (or vice versa), to create a proper gray-box approach is, in my view, the ultimate holy grail quest for space weather forecasting.

\subsection{Relativistic electrons at geosynchronous orbit}\label{sec:GEO}

One of the most challenging tasks in Space Weather forecasting is the prediction of relativistic electrons at geosynchronous Earth orbit (GEO). In particular, it is known that MeV electron fluxes in the Earth's radiation belt are affected by a combination of physical processes that lead to loss and local acceleration \citep{ukhorskiy2012, reeves2013, camporeale2015, baker2018}.
One of the first attempts to use an artificial neural network to predict the flux of energetic electrons at GEO was presented in \citet{stringer1996}, where GOES-7 data was used to forecast the hourly-averaged logarithm of electron flux at energies of 3-5 MeV, one hour ahead.
A feed-forward NN with a single hidden layer was used, varying the number of neurons between 7 and 11. The training set was composed of 1000 hours of data starting from July 1, 1989 and 1000 hours starting from January 1, 1990 were used for testing. The inputs of the NN were the following: 4 hours history of the electron flux, $Kp$ and $D_{st}$ indices, plus the magnetic local time $MLT$ of the predicted electron flux (one hour in the future). Despite achieving very good results in terms of both prediction efficiency and root mean square error ($PE\sim 0.95$ and $RMSE \sim 0.1$) the authors pointed out that due to the strong autocorrelation of Log(flux) at a lag of one hour `... the NN is not much better than linear extrapolation one hour into the future. Indeed [...] to first order, the output depends only on the previous history of the Log(flux)'.
Unfortunately, a comparison against a persistence model (where the output one-hour ahead is simply taken as the value at current time) was not quantitatively performed. \\

Building on this work, \citet{fukata2002} have proposed a more involved NN, known as Elman architecture, which has still only one hidden layer (15 neurons), but it contains feed-back connections from the hidden to the input layer.  They did not use the past history of the Log(flux) as input, but instead they proposed to use the $AL$ magnetic index. History of $AL$ and $D_{st}$ were incorporated in $\sum AE$ and $\sum D_{st}$, that are the summation of the index values from the time of $D_{st}$ minimum in the main phase. They explicitly focused on forecasting one-hour ahead relativistic electron flux during a storm recovery phase. Nine storms in the period 1978-1994 were used for training, and 20 storms for testing. The average value of PE turned out to be 0.71, which is lower than the one reported by \citet{stringer1996} (not calculated on the same test set). They also experimented by dropping out inputs, and noticed that $\sum AE$ is more important than $\sum D_{st}$.\\

A completely different approach has been taken by \citet{o2003}, by leveraging on the expressive power of neural networks as nonlinear regressors combined with a genetic algorithm to systematically explore the 
large dimensional input space. In that paper, the authors explicitly state that their goal was to build a simple empirical model for the energetic electron flux, rather than a forecasting tool. About 700 different NNs were tested, with different combination of outputs and time-lags that included $Kp$, $D_{st}$, $AE$ and ULF wave power, each with time lags ranging from 0 to 48 hours in the past. Interestingly, the best performing (feed-forward) NN used only 5 hidden layer neurons, and four magnetospheric inputs: $D_{st}(t), D_{st}(t-1),D_{st}(t-4)$ and ULF$(t)$. The root-mean-square error on out-of-sample data was 0.122 (the same metric computed for the persistence model was equal to 0.138). The skill score with respect to the persistence model was 22\%. The main goal of that work, however, was to derive an analytical dynamical equation for the time change of the electron flux. Hence, the NN was merely used to identify the most important magnetospheric drivers (solar wind drivers were purposefully excluded). As we will see, this is a recurring theme in the (space) physics community where some sort of dissatisfaction often results by using the black-box machinery of NNs. In that respect, the work of \citet{o2003} was one of the first attempts to open the black-box, deriving an (hopefully easy to interpret) analytical formula, in the context of relativistic electrons dynamics.
The analytical formula was derived using a statistical phase-space analysis technique combined with least squares optimization to fit coefficients. When used for one-hour ahead prediction, the formula achieved a skill score of only 4\%. However, the authors argued that the true value of the dynamic equation was to be appreciated when deriving multiple-hours predictions (with the skill score getting as high as 50\% for 48 hours-ahead prediction). Still, it remains unclear how much of the reported skill scores is due to the goodness of the empirical analytical model, or to the fact that persistence becomes completely useless after a few hours. Finally (as rightly pointed out by the authors) the derived empirical equation has no forecasting value, because it will need future values of $D_{st}$ and ULF wave power to perform multiple-hours ahead predictions.\\

More recent works have focused on developing models for the daily-averaged electron flux (rather than hourly-averaged) \citep{ukhorskiy2004, lyatsky2008, kitamura2011}. From a Space Weather perspective a 1-day ahead forecast is certainly more useful than only a few hours ahead prediction. However, a word of caution is needed, because what 'one-day ahead' really means in most of the papers discussed hereafter is the daily-averaged value of electron flux, that is averaged over a period of 24 hours. By shifting the focus on predicting an averaged quantity, one clearly looses the ability of forecasting sudden large events, that are on the other hand the most interesting and challenging.\\
\citet{ling2010} have systematically tested feed-forward NNs with a single hidden layer, by varying the number of hidden neurons and the number of inputs. They focused on $>$2 MeV electrons measured by the GOES-7 satellite. They used the time history of electron flux and $Kp$ as inputs and tested the best performing NN for a period of 6 months starting from January 1, 2009. A somewhat unsatisfactory results was that the performance metrics seemed to be very sensitive with respect to the size of the training data. The PE for one-day ahead forecast jumped from 0.58 to 0.67 when the training set period was enlarged from 6 months to 1 year. Also, to overcome a neuron saturation problem, the authors settled on a strategy where the model is re-trained daily (with incoming new data), using a training set size of 2 years. In this way, the mean PE for 1, 2, and 3 days forecast is 0.71, 0.49, 0.31, respectively. Finally, the authors reported a better performance (for the period 1998-2008) with respect to the linear filter model REFM (Relativistic Electron Forecast Model) developed by \citet{baker1990}, which is still currently implemented at the NOAA Space Weather Prediction Center (\url{https://www.swpc.noaa.gov/products/relativistic-electron-forecast-model}).\\
The same group of authors have compared their NN model (dubbed FluxPred) against the SWPC-REFM model and the semi-empirical model by \citet{li2001}, for 1,2 and 3 days ahead predictions in the period 1996-2008 \citep{perry2010}. The \citet{li2001} model is a nice example of gray-box modeling, where a physics-based radial diffusion equation is data-augmented, by parameterizing boundary conditions and diffusion coefficients as functions of past solar wind observations. The results of \citet{perry2010} was mostly inconclusive, that is each model did well ad different phases of the solar cycle, and there was no clear winner. Quoting the paper:
 `Over all, the three models give slightly better +1 day and much better +2 day forecasts than persistence [...]. All models are solar cycle-dependent, meaning predictions are better during solar minimum and worse during solar maximum'.
Somewhat hidden in the conclusion of this comparison study, however, lie a suggestion that, almost ten years later, is rapidly becoming a mainstream idea in forecasting, namely the use of ensembles, for example giving different weights to different models during different phases of the solar cycle.\\

Other competing models that are based on more standard statistical analysis are \citet{turner2008} and \citet{kellerman2013}, that reported prediction efficiencies not dissimilar from the models based on NNs. For instance, Figure 8 in \citet{kellerman2013} shows PE as function of time for the period 1992-2010, roughly ranging from 0.4 to 0.8 (1 day ahead) and 0.2 to 0.6 (2 days ahead).\\

Yet another methodology that is complementary to neural networks is the use of an autoregressive model in one of its many ramification. Specifically a NARX (Nonlinear AutoRegressive with eXogenous inputs) model was presented in \citet{wei2011}, where the model performance was specifically compared to the model of \citet{ukhorskiy2004}. Higher average values of PE were reported for the years 1995 and 2000 on which the new algorithm was tested. The extension from NARX to NARMAX (NARX with Moving Average) was presented in \citet{balikhin2011}, and \citet{boynton2013} studied separately several energy bands.
With this approach, an explicit formula linking inputs to output can be derived, from which the long-standing idea of the NARMAX proponents is that some physics insight can be learned (for instance which terms contribute the most). For example, in \citet{wei2011} 30 monomial terms involving solar wind speed $v$, dynamics pressure $P_{dyn}$, $vB_z$ term, $AsyH$ and $Symh$ geomagnetic indices were retained, even though the explicit formula for the forecasting of electron flux was not explicitly given.
\citet{balikhin2016} have compared the performance of the Sheffield SNB$^3$ GEO online forecast tool (based on NARMAX) against the SWPC-REFM model for the period 2012-2014. The accuracy of the forecast in terms of PE was very similar for the two models with SNB$^3$ GEO performing slightly (5\%-10\%) better. Moreover, the authors pointed out that one main deficiency in relativistic electron forecast is the inability of predicting dropouts caused by mangetopause shadowing \citep{turner2012}, which in turns is due to poor forecast of solar wind parameters at L1.

\citet{sakaguchi2013} and \citet{sakaguchi2015} have proposed multivariate autoregressive models based on Kalman filters to forecast GEO and MEO (Medium Earth Orbit) energetic electrons (see also \citet{rigler2004}). A cross-correlation analysis was carried out to identify physical drivers, for a range of time lags, and different $L$ shells. The more highly correlated quantities are solar wind speed, magnetic field, dynamics pressure, and the geomagnetic indices $Kp$ and $AE$.
Predictions from one to 10 days ahead were tested in a 8 months window (September 2012 to December 2013). Interestingly, predictions for GEO yielded smaller prediction efficiency than for $L=3.6, 4.6, 5.6$. Indeed a clear trend was found where orbits closer to Earth (smaller values of $L$) were easier to predict.

\citet{bortnik2018} have proposed a two hidden layers neural network to model radiation belt electrons in the energy range 1.8 - 7.7 MeV and $L<6$, by using the SYM-H index  (sampled at 5 minutes cadence). They used $\sim 188,000$ data points from the Relativistic Electron Proton Telescope (REPT) instrument on-board the two Van Allen Probes, and achieved a correlation coefficient in the range $\sim 0.73-0.84$, generally becoming progressively lower with increasing energy.

All the cited models focused on high energy electrons ($>2$ MeV). One of the very few models that attempted to predict also lower energies has been presented in \citet{shin2016}. They still used a rather simple neural network, although the number of hidden neurons was now increased to 65. Also, it is interesting that their network was designed to forecast simultaneously 24 hourly values of the electron flux in a one day window. Regarding the inputs, a slight novelty with respect to past work was the use of the Akasofu parameter \citep{akasofu1981}.  All input variables were considered with their 4 hours history. The main results were: the prediction efficiency decreases with decreasing electron energy, and it depends on the magnetic local time (more so for low energies than high energies). The reported PE for $>2$ MeV electrons was 0.96 (one hour ahead) and $\sim$ 0.7 (24 hours ahead), when tested with GOES 15 data. However, it has to be pointed out that these metrics have been calculated on the validation test, and not on an independent test set. Hence, the generality of such a good performance was not demonstrated.

Following the current trend in Machine Learning of `going deeper', a deep learning model has finally appeared in the arena of energetic electron flux forecasting, in 2018. The paper by \citet{wei2018} uses a so-called Long Short-Term Memory network (LSTM, \citep{hochreiter1997}), which has been successfully employed in time-series forecasting.  In this paper, both daily and hourly forecast are presented, testing several combinations of inputs and number of hidden neurons (the largest being 512). Three years (2008-2010) were used for testing. 
Maybe because of the computational cost of training a LSTM network, only three inputs were used in all experiments (one of which is always the flux itself). As a result, the prediction efficiency reported is not substantially higher than what was obtained with more traditional networks. For instance, the highest PE for the daily prediction (averaged over one year of forecast) was 0.911.

Finally, the paper by \citet{miyoshi2008} needs to be mentioned, for the simple reason that it appears to be the only model that produces a probabilistic forecast instead of single-point predictions. The importance of probabilistic predictions is a recurring theme in this review paper and they pose an important challenge for future Space Weather research. The model of \citet{miyoshi2008} is not very sophisticated, being based on the statistical analyses of superimposed epochs, taking seasonality and solar wind speed into account. The model is essentially a climatological model, and the forecast is based on long time average (11 years) of observed stream interfaces. Unfortunately, no quantitative metrics were discussed.

\subsection{Recapitulation - Relativistic electrons at geosynchronous orbit} 
Similarly to the predictions of geomagnetic indices, it is hard to draw a straightforward conclusion from the review presented in the previous Section for relativistic electrons at geosynchronous orbit. Many different approaches have been tried, mostly using neural networks, but lessons from past works have not always been taken in consideration. Hence, newer models often did not outperform older ones. Moreover, a trait that undermines most works in the field of Space Weather predictions is the lack of a standard and agreed-on set of `challenges' or benchmarks commonly used to assess and validate different models. As a result, the metrics reported in one paper cannot easily be transferred to another paper, which is trained and tested on different sets. In passing, we note that the space weather community has been involved in the past in community-wide validation efforts, especially to support model transition to operations. One such example concerns the geospace model to predict ground magnetic field perturbations \citep{pulkkinen2013, glocer2016, welling2018}.\\
It appears that the inaccuracies of current models are mostly due to the uncertainties in the forecast of solar wind parameters that are used as drivers to estimate future fluxes of relativistic electrons. Another source of uncertainty might be to the internal magnetospheric dynamics that is not easily captured by black-box models (for instance substorm cycles). As highlighted in \citet{jaynes2015}, a simple causal relationship between a fast solar wind driver and the enhancement of radiation belt electron fluxes might miss the rare occurrences when high-speed solar wind streams do not produce flux enhancements, if the two distinct population of electrons (termed source and seed) are not properly accounted for.\\
We notice that even though most early works have focused on geomagnetic orbit, nowadays we might have enough data to train models that cover a wider range of orbits (with increasing relevance to Space Weather). In this perspective, a gray-box approach can once again be very effective. For instance, the Fokker-Planck (quasi-linear) approach that describes the evolution of particle phase space density through a multi-dimensional diffusion equation \citep{tu2013, drozdov2015} will benefit from a machine learning estimate of boundary conditions \citep{pakhotin2014} and from Bayesian parameterization of diffusion coefficients and electron timeloss \citep{2017AGUFMSM23A2583C}. We emphasize that the model presented in \citet{li2001} represents an early (non-Bayesian) attempt of gray-box modeling, with a large-number of ad-hoc assumptions and empirical chosen parameterization, that could in the future be improved by means of Bayesian data assimilation and machine learning.\\
Finally, most of the considerations about geomagnetic indices predictions (Section \ref{sec:recap_GEO}), hold true for the forecast of relativistic electrons as well. The main challenge in the future will be to extend the predictions to longer time horizon. This will necessarily mean coupling particle forecasts to the forecasts of solar wind conditions, eventually driven by solar images. It will also require to understand and being able to model the propagation of uncertainties from one model to another. 

\subsection{Solar images}
As already mentioned, solar images offer a large amount of information that seems to be well versed for machine learning techniques.
Because the overall amount of data that one would like to use for machine learning can easily exceed hundreds of Gigabytes (SDO produces about 1.5 Tb of data per day), it is important to use some dimensionality reduction techniques.
These are methods that, exploiting the linear or nonlinear relations between attributes in the data, seek to apply a transformation to the original dataset, reducing their initial dimensionality, at the cost of a minimal loss of information. \citet{banda2013} have investigated several dimensionality reduction techniques for large-scale solar images data (linear methods: Principal Component Analysis, Singular Value Decomposition, Factor Analysis, Locality Preserving Projections; and nonlinear methods: Isomap, Kernel PCA, Laplacian Eigenmaps, Locally Linear Embedding). For details on each one of these techniques, the reader is referred to the original publications and references therein.\\
The two tasks where solar images can be effectively used and that we discuss in the following are the prediction of solar flares and of coronal mass ejections propagation time.

\subsubsection{Solar flares}
Most of the works that use solar images tackle the problem of solar flares prediction. Solar flares are a sudden conversion of magnetic energy into particle kinetic energy associated with large fluxes of X-rays. 
Flares are categorized as A-, B-, C-, M-, or X- classes, depending on their X-rays peak flux measured by the Geostationary Operational Environment Satellite (GOES).
Flares forecast is certainly one of the major active area of research in Space Weather, due to their technological consequences, such as radio communication black-outs or increase in satellite drag. 
One of the first attempt to use a neural network to predict flares is probably \citet{fozzard1989}. 17 binary inputs were used to feed a five neurons hidden layer, that resulted in 3 output neurons (one for each flare class: C, M, X).
Another pioneering work was proposed in \citet{borda2002}, where a single hidden layer feed-forward neural network was used to perform a binary classification on the occurrence of flares. They used images from the Argentinian HASTA telescope and selected seven features extracted from the images. The dataset was necessarily small (361 events in total) and they reported an accuracy of 95\% 

More recently, \citet{wang2008} have developed a single hidden layer neural network that uses features based on three quantities extracted from SOHO/MDI images: the maximum horizontal gradient, the length of the neutral line, and the number of singular points of the photospheric magnetic field. Data from 1996-2001 was used for training and the whole year 2002 was used for testing. A full performance analysis was not conducted, but the overall ratio of correct forecast was indicated to be around 69\%. 

\citet{yu2009} have realized the importance of analyzing time sequences of predictors. They used the same three features as in \citet{wang2008} and have employed an analysis based both on autocorrelation and on mutual information, to select the optimal sliding window of past events from which their method would be trained. The chosen window contained 45 data points, with cadence 96 minutes (the sampling intervals of SOHO/MDI magnetograms). They have tested two different machine learning techniques: a Decision Tree, and a Learning Vector Quantization neural network which is a particular version of a NN for supervised classification \citep{kohonen1990}. The main result of the paper was in showing how the sliding window helped in boosting the performance of both methods by about 10\%.

\citet{yu2010} have proposed a method based on a Bayesian network using again the same three features as in \citet{wang2008}.  The Bayesian network is a probabilistic graphical model that connects variables by their respective conditional probabilities. The output of the network is a binary classifier (flare/no-flare), which in this case predicts whether a flare of at least class C1.0 is produced within a 48 hours window. The best model presented in \citet{yu2010} yielded a hit rate of $\sim 88\%$ and Heidke Skill Score $HSS\sim0.7$. 

\citet{bian2013} have investigated a method based on the so-called Extreme Learning Machine (ELM) \citep{huang2006}. ELM have a controversial history, but they can simply be understood as single hidden layer feed-forward neural networks, with the interesting feature of having their hidden weights and biases randomly chosen. The training does not employs a standard iterative algorithm, such as back-propagation or an evolutionary algorithm. Instead, the optimal weights associated to the output layer are calculated via linear algebra (least-square), by pseudo-inverting the matrix associated with the hidden layer. This translates in a much faster training, and performances often competing with standard and deep neural networks \citep{huang2015}. In \citet{bian2013} the total unsigned magnetic flux, the length of the strong-gradient magnetic polarity, and the total magnetic energy dissipation associated to an active region are used as inputs. The prediction method is a combination of an Ordinal Logistic Regression (OLR) method with an ELM. The OLR output consists in 4 probabilities, respectively associated with classes A or B, class C, M, and X. The OLR output is then fed into the ELM to produce a binary classification. The method yielded positive and negative accuracies of about 30\% and 90\% , respectively, for M-class flares.

\citet{boucheron2015} have developed a Support Vector Regression model that predicts the type and the time of the occurrence of a flare. They have extracted 38 spatial features from 594,000 images (time period 2000-2010) from the SOHO/MDI magnetogram. The output of their regression method is a continuous real value, that is then mapped to a given class. They account for the imbalance of the dataset across different classes, by sub-sampling the larger classes (weak flares), and they employ a 100-fold cross-validation strategy. They reported an average error of 0.75 of a GOES class.

\citet{bobra2015} used a support vector machine classifier to forecast whether a region will be flaring or not, by using 13 features (selected among 25 by evaluating their Fisher ranking score \citep{fisher1992}), obtained by the SDO/HMI vector magnetograms. They have identified 303 examples of active regions that have produced flare (either within a 24 or 48 hours window), in the time period May 2010 - May 2014, and 5000 examples of non-flaring active regions. 
They achieve remarkably good results, with the obvious caveat of a limited test set (which is selected as 30\% of the whole dataset, hence resulting in only about 90 positives).
Interestingly, \citet{bobra2015} present an excellent overview of different performance metrics used for binary classification, and some of their fallacies when the dataset is imbalanced, as in solar flare prediction. See also \citet{bloomfield2012} for a discussion of skill scores, in this context.
Previous similar works used line-of-sight magnetic field data, sunspot numbers, McIntosh class and solar radio flux as input attributes \citep{li2007,qahwaji2007, song2009, yuan2010, leka2018}.

\citet{nishizuka2017} have built on the work of \citet{bobra2015} and analyzed the importance of 65 features obtained from 11,700 active regions tracked in the period 2010-2015. The features were obtained from line-of-sight and vector magnetograms (SDO/HMI) and from GOES X-ray data. Moreover, a novelty of this work was to recognize the importance of the chromospheric brightening extracted from the SDO/AIA 1600 \r{A} images. Three machine learning techniques were compared:  k-Nearest Neighbor (k-NN) classifier, a SVM classifier, and an extremely randomized tree (ERT). The k-NN yielded the best results, with a TSS greater than 0.9. A caveat of this work, pointed out by the authors, is that they have used a random shuffle cross-validation strategy, that would artificially enhance performance. They also note that the standardization of attributes strongly affects the prediction accuracy, and that this was not yet widely acknowledged by the solar flare forecasting community. Finally, a somewhat unsettling finding is that the persistent nature of the flares, that is the indication of the maximum X-ray intensity in the last 24 hours turned out the be the most important feature, once again highlighting the importance of persistent models in Space Weather forecasting.

The same authors have presented a model based on a deep neural network in \citet{nishizuka2018}. Here, the fallacy of randomly splitting the training and test sets was openly addressed and rectified. The same features as in \citet{nishizuka2017} were used, with the addition of features extracted from the SDO/AIA 131 \r{A} images, totaling 79 features. The network was designed with 7 hidden layers, each with either 79 or 200 nodes. The output layer produced a two-dimensional vector $(p(M),p(C))$ denoting the probability of a M or C class event, respectively. The final results, tested on the whole 2015 year, were very promising, yielding a $TSS\sim 0.8, 0.6$ for M and C class prediction, respectively.

An important milestone in the use of machine learning for solar flare predictions is represented by the EU-H2020 project FLARECAST, which was explicitly tasked with automatic forecasting based on both statistical and machine learning techniques. A comprehensive report of the project can be found in \citet{florios2018} (see also \citet{massone2018}). Being a fully-dedicated three-years project, there are several aspects worth commenting. All the codes produced in the project have been released and are open-access, thus promising a future legacy and the possibility of a long-standing community-based effort to improve and adopt their methods. \citet{florios2018} presents a detailed comparison between three machine learning methods (a single-layer feed forward neural network, a support vector machine, and a random forest), and some non-machine learning (statistical) methods. They tackle specifically the classification task for $>$M1 and $>$C1 classes, both as a binary and a probabilistic prediction. Overall, seven predictors were chosen (six of which were computed both from line-of-sight and magnetograms and three respective radial component), and several performance metrics were calculated. Interestingly, the paper also provides ROC curves and reliability diagrams for probabilistic forecasts. Although no single method was consistently superior over the whole range of tasks and performance metrics, the random forest was slightly better than the other methods, with the best reported $TSS\sim 0.6$. Also, by using a composite index that weights accuracy, TSS and HSS, and ranking different methods (with different probability thresholds) the random forest scored in the top six positions for both M and C classes forecast. Finally, the paper proves the superior ability of forecasting of the machine learning methods versus the statistical ones. Unfortunately the authors used a random split between training and test sets, which is well known to artificially increase the performance metrics and leaves room for questions about the generalization of the results. 

Several novelties with respect to previous approaches have been introduced by \citet{jonas2018}. They recast the problem from a fixed-time forecast (e.g. 12 or 24 hours ahead prediction) to the prediction of flare occurrence within a certain time window, that is: will an active region produce an M or X class flare within the next T hours? They specifically investigated short-time ($T=2$) and daily ($T=24$) predictions. Similar to \citet{bobra2015}, a strong emphasis was put in the imbalanced nature of data (with a positive/negative ratio of 1/53 for the 24-hour prediction). They appropriately split the data into training and test sets, by segregating all the data associated with the same active region to either one of the sets. One of the most interesting novelty, from a machine learning perspective, is that, along with the classical features derived from vector magnetic field (same as in \citet{bobra2015}), and features that characterize the time history of an active region, they also considered features automatically extracted from HMI and AIA image data. They did that by applying a filtering (convolution) procedure, followed by a non-linear activation function and down-sampling the filtered image to a single scalar. In principle, this procedure is not very dissimilar to what is done in a convolutional neural network (CNN), except the filters are not trained to minimize a specific cost function, but they are chosen a priori.\\
This is an interesting approach that allows to compare the predictive power of physics motivated and automatically extracted features. Despite having automatically generated features from 5.5 Tb of image data, taken between May 2010 and May 2014, and to have at their disposal a rich set of features, the authors have then resorted to use linear classifiers. Two methods were compared (with different regularization term), both designed to minimize the TSS. They found that the (automatically generated) photospheric vector-magnetogram data combined with flaring history yields the best performance, even though by  substituting the automatically generated features with the physical ones does not strongly degrades the performance (within error bars, $TSS\sim 0.8$). Somewhat surprisingly, when using all combined features (physics-based, flare history, and automatically generated from HMI and AIA) the performance was appreciably lower than in the previous two cases. In conclusion, as pointed out by the authors, the results of this paper were only slightly better than the original results presented in \citet{bobra2015}. Yet, it would be interesting to asses if the automatically generated features would benefit more from a non-linear classifier. 

It is interesting to notice that all the works commented above do not use solar images directly as inputs for the classifiers, but instead they rely on features extracted from the solar images. The majority of the models use features that have an interpretable physical meaning. In this sense it seems that the solar flare forecasting community (even its machine learning enthusiast portion) has not yet embraced a full black-box approach where the feature extraction is fully automated. 

The single exception is represented by the recent paper by \citet{huang2018}. Here, images of active region patches of size $100\times 100$ pixels, extracted both from SOHO/MDI and SDO/HMI, are directly fed into a convolutional neural network, without a prior hand-crafted feature extraction. Two convolutional layers with 64 $11\times 11$ filters each are used. As it is customary, the features extracted from the convolutional layers are then fed into two fully connected layers, respectively with 200 and 20 neurons, that finally produce a binary output. The model forecasts C, M and X class flares for 6, 12, 24, and 48 hours periods. The performance metrics do not seem to yield superior results than early works with pre-chosen features. The TSS ranges between $\sim 0.5$ for C class and $\sim 0.7$ for X class.

\subsubsection{Coronal Mass Ejections and solar wind speed}
Coronal Mass Ejections (CME) are violent eruptions of magnetized plasma that leave the surface of the Sun with speed as large as 1,000 km/s. Predicting the evolution of a CME as it expands away from the Sun and travels towards Earth is one of the major challenge of Space Weather forecasting. Indeed, it is well known that the speed and the magnetic field amplitude and orientation of the plasma that impinges on the Earth's magnetosphere are causally related to the onset of geomagnetic storms \citep{gosling93}. The low density magnetized plasma that constitutes the solar wind is well described by magneto-hydrodynamics (MHD), and the standard way of forecasting CME propagation adopted by all major space weather forecasting providers, is to resolve numerically the MHD equations, with boundary and initial conditions appropriate to mimic an incoming CME~\citep[see, e.g.,][]{parsons2011, lee2013,liu2013, scolini2018}. We note in passing that the problem of determining boundary and initial conditions (which are not completely observable) constitute a core challenge for quantifying the uncertainties associated with numerical simulations \citep{kay2018}, and where machine learning techniques can also be successfully employed, especially within the gray-box paradigm commented in Section \ref{sec:ML_SW}.

Because many models and codes have been developed in years by different groups, an effort to collect and compare results of different models is being coordinated by the NASA's Community Coordinated Modeling Center (CCMC), with a public scoreboard available at \url{https://kauai.ccmc.gsfc.nasa.gov/CMEscoreboard/}.
The web-based submission form allows any registered user to submit in real-time their forecast. \citet{riley2018} have recently presented a statistical analysis of the 32 distinct models that have been submitted in the interval 2013-2017, for a total of 139 forecasts. Even though different teams have made different number of submissions (ranging from 114 forecasts from the NASA Goddard Space Weather Research Center to just 1 from the CAT-PUMA team), this paper provides a useful baseline against which any new model should compare its performance. We refer the reader to the original paper to appreciate the many caveats of the statistical analysis (for instance, the bias due to choosing which events to submit), but for the purpose of this review it is sufficient to capture the overall picture. The mean absolute error of the arrival time averaged over models ranges between $MAE=11.2$ (2013) and $MAE=22.6$ (2018) hours, with typical standard deviations of $\pm$ 20 hours. Interestingly, the authors noted the somewhat discouraging result that forecasts have not substantially improved in six years.

\citet{liu2018} have presented a model to predict the arrival time of a CME, using Support Vector Machine. A list of 182 geo-effective CMEs was selected in the period 1996-2015, with average speeds ranging between 400 and 1500 km/s. Eighteen features were extracted both from coronagraph images (SOHO/LASCO) and from near-Earth measurement (OMNI2 database). By ranking the importance of the features, based on their Fisher score, they showed that the CME average and final speed estimated from the Field of View of LASCO C2 are by far the most informative inputs, followed by the CME angular width and mass, and the ambient solar wind Bz. The performance of the method was remarkable, with a root mean square error $RMSE\sim 7.3$ hours. 

The relationship between CMEs and flares is still not completely understood. Indeed, some active regions trigger both a flare and a CME, while {in other regions} flares are not associated to a CME. In \citet{bobra2016}, the authors have developed a classifier based on Support Vector Machine to study features that can distinguish between the two cases and eventually to forecast whether an active region can produce an M or X class flare. The methodology is very similar to the one in \citet{bobra2015}, with 19 physically motivated features extracted from the SDO/HMI vector magnetometer. The best performing method yields a $TSS\sim 0.7$ and uses no more than six features. 

\citet{inceoglu2018} have extended the methodology presented in \citet{bobra2015} devising a three categories classifier: the new method predicts if an active region will produce only a flare, a flare associated with CME and solar energetic particles (SEP), or only a CME. The machine learning algorithms explored are a (multi-class) Support Vector Machine, and a single hidden layer Neural Network. The work builds on the previous findings of \citet{bobra2015} in choosing the features and selecting active regions from SDO/HMI images. Several models were built and compared with prediction times ranging from 12 to 120 hours. The performance in terms of TSS was very high, with the best models achieving $TSS\sim0.9$. 

The study of CME propagation is obviously only a part of the bigger challenge of being able to accurately model and forecast the ambient solar wind properties, in particular speed and magnetic field. A comprehensive review about the state-of-the-art in solar wind modeling resulting from a workshop on `Assessing Space Weather Understanding and Applications' can be found in \citet{macneice2018}. One of the main conclusions of the review is that currently empirical models outperform both semi-empirical and physics-based models in forecasting solar wind speed at L1, and all models perform poorly in forecasting $B_z$.

One of the main application of machine learning in forecasting solar wind speed 3 days ahead was presented in \citet{wintoft1997,wintoft1999}. A potential field model was employed to expand the photospheric magnetic field obtained from magnetograms to 2.5 $R_s$. A time series of the source surface magnetic field was then fed to a radial basis neural network to output the daily average solar wind speed. The best model gave a $RMSE\sim 90$ km/s and $cc\sim 0.58$.

The hourly averaged solar wind speed was predicted using Support Vector Regression (SVR) in \citet{liu2011}. Several case studies were presented focusing either on CME arrival or coronal hole high-speed streams, but overall a certain degree of one-step persistence seemed to dominate the results. Indeed, the fact that a persistence model yields an excellent performance in short-term predictions has been known for long. This has to do with the fact that solar wind variations occur on average on long time scales and that sudden variations are relatively rare. Hence, when averaged over long time periods the performance calculated by means of simple metrics such as $RMSE$ is not sensitive to large errors in predicting sudden changes of speed.

A simple statistical model (not machine learning) based on the construction of conditional probability density functions (PDF) has been presented in \citet{bussy2014} and later refined in \citet{bussy2016}. The PDF model is based on past speed values and slope (i.e. if the speed is increasing or decreasing) and it outputs a probabilistic prediction by linearly combining the prediction based on the PDF and the actual speed observed one solar rotation ago. The PDF model was shown to perform equal or better than the persistence model for all times up to 5 day prediction (the further out the prediction, the better the model), with an error ranging from $RMSE\sim 66$ to $RMSE\sim 90$ km/s.

Inspired by the model of \citet{wintoft1997}, \citet{yang2018} have developed a neural network-based model that predicts solar wind speed 4 days in advance. The Potential Field Source Surface (PFSS) model was used to derive 7 attributes, to which they added the solar wind speed 27 days in the past. Once again, a persistence model provides a very strong baseline. Indeed, a prediction based solely on the past solar wind speed (approximately one solar rotation in the past), yields already a correlation coefficient $cc\sim0.5$ and a $RMSE\sim 95$ km/s. The final model results in $cc\sim0.74$ and $RMSE\sim 68$ km/s, which is probably the state of the art, as of today.

Other works that have tackled the problem of solar wind velocity predictions are \citet{liu2011, innocenti2011, dolenko2007}.

\subsection{Recapitulation - Solar Images}
The first thing that appears evident by reviewing the literature of machine learning techniques applied to to forecast of solar flares, coronal mass ejections and solar wind prediction is that solar images are rarely used directly as inputs. Indeed, with the exception of 
\citet{huang2018}, all the presented works use solar images (magnetograms and extreme ultra violet (EUV) images) to extract features that are either hand-crafted (physics-based), or automatically extracted via pre-defined filters. 
One might wonder whether this choice is simply dictated by the computational cost of processing images and having a large dimensional input in machine learning algorithms. As highlighted by the FLARECAST project \citep{florios2018} machine learning techniques have been shown to give better performance than statistical methods. This motivates the quest for more advanced and accurate techniques. The three problems discussed in the last Section, however, are profoundly different in nature. The imbalanced nature of solar flares data makes it hard to judge the generality of the results. In this respect, it has to be noticed that almost exclusively SDO images have been used. Despite the wealth of information and the high resolution provided by SDO, an open question remains of whether 8 years of data (that is, less then a solar cycle) are adequate to train, validate and test a machine learning model. They are probably not, and it will be worth to try combining SDO and SOHO images to have a larger dataset. This is not straightforward, since the instruments are different, and it would require some careful preprocessing.
Regarding CMEs propagation and solar wind speed forecast, it seems that simple empirical models are still hard to beat and that adding complexity in terms of machine learning algorithms often does not pay off. However, it is also true that advanced (computationally demanding) machine learning techniques, such as deep learning, have not been tried yet. This certainly seems to be a field where the combination of physics-based models, such as MHD propagation simulations, and machine learning models might be successfully integrated in a gray-box approach.

\subsection{Other space weather related areas}
There are several other areas where machine learning has been applied in a less systematic way, but that are nonetheless promising for a data-driven approach.
Plasmaspheric electron density estimation has been proposed in \citet{zhelavskaya2017, zhelavskaya2018}. Concerning the ionosphere-termosphere region, ionospheric scintillation has been modeled in \citet{mcgranaghan2018, linty2019, rezende2010, lima2015, jiao2017}. The estimation of maps of Total Electron Content (TEC) has been tackled in \citet{habarulema2007,watthanasangmechai2012, habarulema2009, wintoft2000, hernandez1997, acharya2011, leandro2007, tulunay2006}. The foF2 parameter (which is the highest frequency which reflects from the ionospheric F2-layer)has been studied in \citet{wang2013, poole2000, oyeyemi2005}, and termosphere density in \citet{perez2014,choury2013}.

\section{New trends in machine learning}\label{sec:new_trends}
A somewhat different interpretation of machine learning with respect to what has been discussed until now, divides its applications into two fields. On one side, machine learning can be used to accelerate and automate a number of tasks, that are very well understood and mastered by human intelligence.
Supervised classification is a typical example, where the advantage of `teaching' a machine how to distinguish objects stays in the ability of classifying them in an automatic, faster and possibly more accurate way than it would be done by humans.
On the other side, machine learning can be used for \emph{knowledge discovery}, that is to truly deepen our understanding of a given system, by uncovering relationships and patterns not readily identifiable. A remarkable example is in algorithms learning how to play games without knowledge of any pre-programmed rule, using techniques that belong to a sub-field of machine learning called reinforcement learning (RL), which is orthogonal with respect to what has been discussed in Section \ref{sec:ML_SW}. A reference textbook is \citet{sutton2018}. The most famous example is now AlphaGO, that has defeated Lee Sedol, the world-champion in the game of Go. This might not sound so extraordinary (particularly to non Go players, like myself). After all it was already clear in 1997, with the defeat of Chess-master Kasparov from DeepBlue (IBM), that computers could beat human masters in complex games (although it has to be noted that DeepBlue and AlphaGO are technically very different, with the latter not being specifically pre-programmed). However, what has happened in the AlphaGo-Seidol game was something that will stay in the annals of artificial intelligence. The computer played (at least one time) a move that was simply not understood by the experts. It was at first believed to be a mistake, until it became clear that the software had actually discovered a new strategy, that the collective intelligence accumulated in thousands of years of playing had not yet considered. This is knowledge discovery at its finest (see \citet{metz2016, holcomb2018} for an account of the now famous Move 37).\\
Obviously many applications live in between the two fields of discovery and automation, and machine learning is moving at such a fast pace that more and more applications and ideas will be unveiled in the coming decade. In this Section I describe three new ideas in machine learning that I believe will soon become tools for scientific discovery in physics.\\

{\bf Physics-informed neural networks}. We have described how a gray-box approach combines data-driven machine learning with physics-based simulations (see Section \ref{sec:ML_SW}). The field of scientific computing, that is the ability of numerically solving equations, is the backbone of numerical simulations. It has solid roots in half a century of discoveries in computer science and in the even longer history of numerical analysis. As such, it is a discipline that, so far, seems to be immune to machine learning. However, recent works have investigated how to solve differential equations by using deep neural networks \citep[see, e.g.,][]{rudy2017, raissi2018}.  The underlying idea is that a neural network constructs a nonlinear map between inputs and outputs that, as complex as it might be, is analytically differentiable. Hence, one can enforce a set of equations to be very accurately satisfied on a given number of points in a computational domain. This idea does not differ very much from mesh-less grid methods, that expand the function of interest into a basis (for instance, using radial basis functions) \citep[see e.g., ][]{liu2002, fasshauer1996}.
The main difference resides in the fact that neural networks offer a much richer set of basis, in terms of functions that can be represented. Examples have been shown where fluid equations, such as the Burgers equation, can be solved accurately, even reproducing shocks \citep{raissi2017}, and free parameters be estimated from data \citep{raissi2017b}.
Being able to solve partial differential equations with machine learning probably does not exclude the need to solve the same equations with standard methods, and the two approaches need to be understood as complementary. However, it is worth investigating in which situations an expensive physics simulations (for instance the MHD expansion of the solar wind), might be substituted by a quicker machine learning approximation.\\

{\bf Automatic machine learning}. There is a certain dichotomy in essentially all the neural network works commented in this review. 
While on one hand, by resorting to neural networks, one surrenders any hope to describe the problem at hand by means of a clear, intelligible input-output relationship (and the use of a black-box machinery is indeed an abundant criticism), on the other hand it still seems that the typical work does not exploit in full the capability of neural networks, by resorting to the most simple architecture, the multi-layer feed-forward network. In a sense, a certain degree of understanding how the network works and the ability to grasp it graphically is still preserved. In passing, the reader might have noticed that I have (intentionally) not included here the typical graph of a NN. Such a visual explanation of neural networks can be found in the majority of papers in this review. \\
Of course, the main reason to use \emph{simple} networks might simply be the computational cost of training and comparing different architectures. Still, from the perspective of seeking the best nonlinear map that describes data using a neural network, there are no particular reasons to stick to a simple, human-intelligible network. Based on this premise, a recent trend called \emph{auto-ML} goes in the direction of automatically search for the most performing architecture, and to optimize a certain number of hyper-parameters. From a mathematical perspective, this is again an optimization problem, even though the search space is now discrete (e.g., number of neurons). Hence, promising techniques use genetic algorithm to make different networks compete, in search of the most performing one for a given task \citep{hutter2019}.\\ 
In the field of space weather, auto-ML might be particularly effective when dealing with different subsystems, such as the radiation belts, the ring current, the solar wind, etc., that have both internal dynamics and external interactions between them. Being able to explore the most efficient graph connections amongst neurons pertaining to different physical domains might result in a better ability of encoding the complex Sun-Earth interactions.

{\bf Adversarial training}. A major weakness of neural networks is that they are vulnerable to adversarial examples. In the context of image classification, for example, an adversarial example is an image that has been produced by applying a small perturbation to an original image. That perturbation can be tailored in such a way that causes the algorithm to mis-classify the image. A straightforward way of generating adversarial examples has been proposed in \citet{goodfellow2015}. If we denote with $\mathbf{x}$, $y$, and $L(\mathbf{x},y)$ the original input, the target output, and the loss function, respectively, then a new input 
\begin{linenomath*}
\begin{equation}
 \mathbf{x}' = \mathbf{x} + \varepsilon\sign\left(\nabla_{\mathbf{x}} L(\mathbf{x},y) \right)
\end{equation}
\end{linenomath*}
(where $\varepsilon$ is a small value) will result in a larger loss function than the one calculated on the original input $\mathbf{x}$. Simply put, the adversarial example perturbs the input in the `right' direction to increase the loss. Taking into account adversarial examples makes a machine learning model more robust and generalizable.\\
An important application of the idea of adversarial examples are Generative Adversarial Networks (GANs) that can be used to artificially generate inputs hence augmenting data or filling gaps in the data. A recent example of the use of GANs in space physics is the generation of total electron content (TEC) maps \citep{chen2019}.

\section{Conclusions}\label{sec:conclusions}
More than a decade ago, in a review paper of that time, \citet{lundstedt2005} pointed out that physics-based model were under development, but that it could have taken as long as 10 years for those models to really be useful for forecasting.
 The prediction was spot on, as only recently forecasters have started to use more systematically global simulations to forecast geomagnetic activity \cite[see e.g.,][]{kitamura2008, pulkkinen2013, welling2017, liemohn2018}.
On the other hand, early adopters of machine learning (even before the term was widely used) have encouraged the physics community to look more closely at machine learning techniques, also at least a decade ago. For instance, \citet{karimabadi2007} have prototyped a machine learning technique to automatic discover features such as flux transfer events \citep{karimabadi2009}.\\
Figure \ref{fig:hist} suggests that the field has now reached some degree of recognition within the space physics and space weather community. Forecasting based on machine learning techniques is certainly not yet the mainstream approach, but there is no reason to doubt that it will become more and more predominant within the next decade. My personal prediction is that in particular the gray-box approach, that I have tried to highlight and comment several times in this review, will slowly take the place of more conventional physics-based models.\\ 

A certain skepticism surrounding the use of machine learning in physics is undeniable. The main argument revolves around the fact that we (supposedly) do not still understand why certain machine learning techniques work, and this is in stark contrast to our perfect understanding of physics laws (Newton's, Navier-Stokes, Maxwell's, etc) and their assumptions and limitations. 
{In reality, physics-based models fail at least as often as empirical models in space weather forecasting, for the simple reasons that their assumptions can usually be checked only a-posteriori and that they still rely on several empirical (data-derived) parameterizations.}\\
This review is definitely not the place where to discuss in length one or the other thesis. However, I would like to briefly mention that research on the mathematical foundations of machine learning and its connection with physics is a growing and intense area. The reader interested in the theme of why machine learning works so well in physics and why deep learning often works better than shallow learning should consult, e.g., \citet{lin2017,poggio2017}.\\

Going back to the field of Space Weather predictions, I would like to conclude with a list of challenges that I envision will be tackled within the next decade, and that I encourage the community to address. Whether or not this research will result in better forecasting capabilities is hard to say, but I am pretty confident that it will at least result in a better understanding and acquired knowledge of the Sun-Earth system.

\subsection{Future challenges in machine learning and Space Weather}

{\bf The information problem}. What is the minimal physical information required to make a forecast? This problem lies at the heart of the failure or success of any machine learning application. If the features chosen as input do not contain enough information to setup the forecasting problem as physically meaningful in terms of cause-effect, the machine learning task is hopeless. Even though our understanding of the underlying physics of most space weather problems can help formulating a well-posed task, this remains an open challenge in many applications. For instance, is it sufficient to use solar images from magnetograms and EUV channels to be able to predict solar flares? The approach that uses tools from information theory should help answer some of this questions, even if they provide rather qualitative indications.

{\bf The gray-box problem}. What is the best way to make an optimal use of both our physical understanding, and our large amount of data in the Sun-Earth system? The models that are routinely used in space weather forecasting are inevitably approximated, and rely on the specification of several parameters that are often not observable. An example is the diffusion coefficients in the quasi-linear approach for the Earth's radiation belts. 
An appropriate popular aphorism in statistics is that \emph{all models are wrong, some are useful} \citep{box1979}. The physics-based models employed in predicting solar wind propagation and CME arrival time are not competitive with respect to empirical models \citep{riley2018}. How do we incorporate a gray-box approach in space weather modeling? Learning from other geophysical fields, promising approaches seem to be Bayesian data assimilation and parameter estimation. In turn, these approaches open the problem of running ensemble simulations in a reasonable amount of time, which results in the surrogate problem (see below).
On the other hand, non-Bayesian approaches to solve an inverse problem, based on deep learning might be equally promising.

{\bf The surrogate problem}. What components in the space weather chain can be replaced by an approximated black-box surrogate model? What is an acceptable trade-off between lost of accuracy and speed-up? For instance, in scientific computing and uncertainty quantification, several methods have been devised to combine a few high-accuracy simulations with many low-accuracy ones to quickly scan the space of non-observable input parameters. These methods take the name of multi-fidelity models \citep{fernandez2016, forrester2007}. On the other hand, is it possible to devise surrogate models that enforce physical constraints, such as conservation laws, hence reducing the search space of allowed solutions?

{\bf The uncertainty problem}. Most space weather services provide forecast in terms of single-point predictions. There is a clear need of understanding and assessing the uncertainty associated to these predictions. Propagating uncertainties through the space weather chain, from solar images, to L1 measurements, to magnetospheric and ground-based observations is a complex task that is computationally demanding. The UQ community has devised methods to estimate uncertainties in ways that are cheaper than brute-force, and the space weather community should become well-versed in these techniques. The mainstream approach is called \emph{non-intrusive}, and it boils down to collecting an ensemble of runs using a deterministic model and estimating uncertainties from the statistics of the ensemble. The two difficulties of this approach (that is essentially a Monte Carlo method) are in selecting how to scan the input parameter space to produce the ensemble, and how to estimate the probability associated with each individual input parameter. More details on these problems can be found in \citet{camporeale2019}.

{\bf The too often too quiet problem}. Space weather dataset are typically imbalanced: many days of quiet conditions and a few hours of storms. This poses a serious problem in any machine learning algorithm that tries to find patterns in the data. It is also problematic for defining meaningful metrics that actually assess the ability of a model to predict interesting events. On one hand, the problem will automatically alleviate with more and more data being used for machine learning. On the other hand, it raises the question about whether it is appropriate to augment the available data with synthetic data that hopeful do not degrade the information content of the dataset. Something that will be worth pursuing in the future is to use simulation data in the machine learning pipeline.

{\bf The knowledge discovery problem}. Finally, the problem that many physicists care the most when thinking about using machine learning. How do we distill some knowledge from a machine learning model, and improve our understanding of a given system? How do we open the black-box and reverse-engineering a machine learning algorithm? As already mentioned, this is now a very active area of research in computer science and neuroscience departments. Ultimately, a machine learning user is faced with the problem of focusing either on the \emph{make it work}, or on the \emph{make it understandable}. 
I believe that this is a dilemma too well known to space weather scientists, being a discipline rooted in physics but with a clear operational goal. {I also think that a systematic machine learning approach to space weather will, in the long term, benefit both the forecasting and the science behind it.}\\
In conclusion, the argument behind the push of better understanding what is going on in the black-box is simple: how can we trust an algorithm that we do not have full control of? However, as pointed out from Pierre Baldi, we trust our brain all the time, yet we have very little understanding of how it works \citep{castelvecchi2016}.

\acknowledgments
This work was partially supported by NWO Vidi grant 639.072.716. This project has received funding from the European Union’s Horizon 2020 research and innovation programme under grant agreement No 776262 (AIDA). No data was used.


%
 \bibliography{camporeale_bib}
%




\end{document}